\documentclass[aps,prd,twocolumn,nofootinbib,amsmath,amssymb]{revtex4-1}
\usepackage{CJK}                     
\usepackage[dvips]{graphicx}         
\usepackage{mathptmx}                
\usepackage{bm}
\usepackage{physics}
\usepackage{dcolumn}                 
\usepackage{multirow}                
\usepackage{xcolor}
\usepackage{hyperref}
\hypersetup{
breaklinks=true,colorlinks=true,
linkcolor=blue,citecolor=blue,filecolor=magenta,urlcolor=black}

\def\be{\begin{equation}} \def\ee{\end{equation}}
\def\bal#1\eal{\begin{align}#1\end{align}}
\def\non{\nonumber}
\def\eps{\varepsilon}
\def\phi{\varphi}
\def\la{\lambda}
\def\al{\alpha}
\def\om{\omega}
\def\ms{\,M_\odot}

\def\km{\;\text{km}}
\def\fm3{\;\text{fm}^{-3}}
\def\xp{x_p}
\def\khz{\;\text{kHz}}
\def\taues{\tau_\text{GW}}
\def\mpns{M_{\text{PNS}}}
\def\rpns{R_{\text{PNS}}}
\def\mmax{M_\text{max}}
\def\smax{s_1} \def\smin{s_0}

\begin{document}

\title{$f$-mode oscillations of protoneutron stars}

\begin{CJK*}{UTF8}{gbsn}

\author{Zi-Yue Zheng (郑子岳)$^{1}$}
\author{Ting-Ting Sun (孙婷婷)$^{2}$}
\author{\hbox{Huan Chen (陈欢)}$^{2}$}
\author{\hbox{Jin-Biao Wei (魏金标)}$^{2}$}
\author{Xiao-Ping Zheng (郑小平)$^{1,3}$}\email{Email:zhxp@ccnu.edu.cn}
\author{G. F. Burgio$^{4}$}
\author{H.-J. Schulze$^{4}$}

\affiliation{
$^{1}$Institute of Astrophysics, Central China Normal University,
Luoyu Road 152, 430079 Wuhan, China\\
$^{2}$School of Mathematics and Physics, China University of Geosciences,
Lumo Road 388, 430074 Wuhan, China\\
$^{3}$School of Physics, Huazhong University of Science and Technology,
Luoyu Road 1037, 430074 Wuhan, China\\
$^{4}$INFN Sezione di Catania, Dipartimento di Fisica,
Universit\'a di Catania, Via Santa Sofia 64, 95123 Catania, Italy
}

\begin{abstract}
We investigate nonradial $f$-mode oscillations
of protoneutron stars in full general relativity,
employing equations of state described
by the Brueckner-Hartree-Fock theory or the relativistic mean field model,
while assuming isentropy and fixed lepton fractions for the internal structure.
The validity of various universal relations for cold neutron stars
involving $f$-mode characteristics and macroscopic properties of the star
is confirmed for those isentropic protoneutron stars.
Prospects of observations are also discussed.
According to simulation results,
we then model details of the thermal and trapping profiles
in a PNS with the canonical mass.
The corresponding $f$-mode frequencies
and gravitational-wave strain amplitudes are presented.
The validity of the universal relations
during the evolution to the formation of a cold neutron star is confirmed.
\end{abstract}

\maketitle
\end{CJK*}

\section{Introduction}

Neutron stars (NSs) are the densest stars observed in the Universe
and provide natural laboratories
for studying high-density nuclear matter (NM).
The interior of a NS can reach several times the nuclear saturation density
$\rho_0 \approx 0.16\fm3$.
Theoretically,
the interior of the NS is characterized by the equation of state (EOS)
that links pressure and density.
The EOS connects the microscopic properties of NM inside the star
to the macroscopic structure of the NS,
such as mass and radius.
Currently,
the high density in the core of NSs
presents a major challenge in creating a nuclear EOS for NSs
that accurately describes the behavior of dense matter
under various conditions,
due to the lack of exact computational methods
for nonperturbative strong interactions.

In the studies of cold NSs,
the impact of temperature on the EOS is often neglected.
However,
during some intense dynamic processes
such as core-collapse supernovae (CCSNe)
\cite{Constantinou14,Janka16},
and binary NS mergers (BNSMs) \cite{Figura20,Figura21},
temperature may play an important role in dynamic evolution.
In general,
when a massiv-star progenitor
reaches the effective Chandrasekhar mass,
gravity causes its core to collapse to nuclear densities,
resulting in a protoneutron star (PNS)
\cite{Bethe90,Prakash97,Jank12,Burrows13}.
These so-called CCSN events are expected to
be a significant source of gravitational wave (GW) radiation.
PNSs are extremely hot and lepton-rich,
leading to the trapping of abundantly produced neutrinos
in this extreme circumstance \cite{Pons99,Nicotra06}.
Therefore,
the EOS for PNSs,
where one must consider the dependence
on lepton fraction and temperature,
is quite different from that of cold NS matter
\cite{Barranco81,Chen14,Lu19,Shen20,Wei21,Liu22}.
Investigations of PNSs and associated phenomena offer
valuable insights into the properties of NSs
and a deeper understanding of hot and dense NM,
which are difficult to acquire on Earth.

Following the first direct observation of GWs
from a binary black hole (BH) merger \cite{Abbott16},
more GW signals have been detected,
including those from BNSMs \cite{Abbott17a,Abbott20a}
and the NS-BH merger \cite{Abbott19a}.
These observations have opened a new avenue
for investigating the internal structure of NSs,
enhancing our understanding of dense NM \cite{Abbott18}
and astrophysical processes under extreme conditions \cite{Abbott17b}.
For example,
GWs from the inspiral phase have already
set constraints on the EOS of NM at zero temperature,
showing consistency with radii and tidal deformabilities
\cite{Abbott18,Miller21,Pang21,Raaijmakers21,Rutherford24}.
Moreover,
the temperature dependence of EOSs is crucial
for modeling the structure and evolution of NSs
during BNSMs \cite{Baiotti19,Figura21}.
Thus,
future detection of a postmerger GW signal
could yield insights into the EOS of hot and dense NM
\cite{Bauswein11,Rezzolla16}.
Some studies suggest
the remnant of a BNSM may be a stable NS
\cite{Stergioulas11,Bauswein15,Baiotti16}
or a hypermassive NS existing for a short duration
before collapsing into a BH \cite{Baiotti16,Gill19,Figura21}.
Consequently,
investigating PNSs can also offer important insights into
the remnants of a BNSM.

Stellar oscillations serve as a sensitive probe of
the internal composition of NSs
and may offer valuable insights into the challenges
posed by high-density NM.
Notably, the nonradial oscillations (NROs)
with an order of the spherical-harmonic decomposition $l\geq2$
can convey information about the core of a NS
through GW emission \cite{Thorne67},
which may occur during a CCSN \cite{Radice19},
or during the post-merger phase of a BNSM
\cite{Kokkotas01,Stergioulas11,Vretinaris20,Soultanis22},
or in an isolated perturbed NS \cite{Doneva13}.
For a nonrotating star,
NROs can be classified into several distinct modes.
Among these, the fluid modes include the $p$ (pressure) mode,
the $f$ (fundamental) mode, and the $g$ (gravity) mode,
each characterized by its dominant restoring force governing the perturbations.
Buoyancy acts as the restoring force for the $g$ mode,
bringing perturbed fluid elements back into equilibrium.
In contrast, the restoring forces
for the $f$ mode and $p$ mode are primarily pressure-driven.

For cold NSs,
the $^2g_1$-mode eigenfrequencies are relatively small
and serve as effective probes for exploring
new degrees of freedom inside the star,
such as deconfined quark matter
\cite{Weiwei20,Bai21,Jaikumar21,Constantinou21,Zhao22a,Zheng23},
hyperons \cite{Dommes16,Tran23},
and superfluidity \cite{Kantor14,Hang16,Dommes16,Rau18,Burgio21}.
In contrast,
the $p$-mode eigenfrequencies are too high to be observed \cite{Kunjipurayil22},
exceeding the sensitivity range of next-generation GW detectors.
The fundamental $f$-mode eigenfrequencies of cold NSs,
$f_f = \Re\om_f / 2\pi \sim 1.3-3 \khz$
lie between those of $g$ and $p$ modes.
It was argued that the $f$ modes can be strongly excited
by the spin and eccentricity during the inspiral phase of BNSMs
\cite{Chirenti17,Steinhoff21}
or by the glitching behavior of an isolated star
\cite{Sidery10,Ho20,Haskell24}.

Andersson and Kokkotas \cite{Andersson98,Kokkotas99}
first connected the
$f$-mode dimensionless frequency or dimensionless damping time
to the mean mass density
or compactness parameter $\beta=M/R$ of the cold NSs.
Therefore, one might estimate the macroscopic properties of NSs,
such as mass and radius,
by correlating these quantities with the measured frequency
and damping time of a particular oscillation mode.
This technique is known as asteroseismology,
which could help us infer NS features from nonradial oscillations.
Subsequently \cite{Lau09,Chirenti15},
a much more precise EOS-insensitive universal relation (UR) for cold NSs
between dimensionless frequency $\Omega_f \equiv M\om_f$
and dimensionless moment of inertia (MOI) $\bar{I} \equiv I/M^3$
was found.
Furthermore,
as analyzed in \cite{Chan14,Sotani21,Zhao22b},
the $\Omega_f-\bar{I}$ UR implies a related $\Omega_f-\Lambda$ UR,
since the EOS-insensitive $\bar{I}-\Lambda-\bar{Q}$ UR is well known
\cite{Yagi13}.
These URs have been examined in cold NSs with various configurations,
including purely hadronic stars \cite{Sotani21},
hyperonic stars \cite{Pradhan22},
hybrid stars with Maxwell construction \cite{Zhao22b}
or quark-hadron crossover \cite{Sotani23}
or hadron-quark pasta structure \cite{Zheng25},
and strange-quark stars \cite{Zhao22b}.

On the other hand,
recent advancements in numerical simulations of CCSNe
\cite{Andresen17,Richers17,Takiwaki18,Andresen19,Torres-Forne19,Vartanyan20,
Andersen21,Bizouard21,Vartanyan23,Jakobus23,Bruel23,Afle23,Sotani24}
have led to significant contributions in
enhancing our understanding of GW asteroseismology for PNSs.
So far,
studies has identified the dominant frequency in GW signals from CCSNe
as the $f$-mode oscillations
\cite{Morozova18,Sotani19,Sotani20,Wolfe23},
while the signals in the very early phase just after core bounce
are attributed to $g$-mode oscillations
\cite{Sotani20,Wolfe23}.
Meanwhile, the simulations indicate that the major part of GW energy
is concentrated in the lowest-order eigenmodes,
particularly $^2g_1$ and $^2f$ modes \cite{Torres-Forne19}.

Currently,
the application of asteroseismology to PNS studies has been growing,
providing crucial insights into the connection
between GW astronomy and PNS macroscopic properties.
According to the literature
\cite{Torres-Forne19,Sotani19,Sotani20,Sotani21,Bizouard21},
within $\sim 1\,$s of the postbounce phase,
the frequencies of $g$-mode oscillations
depend primarily on the surface gravity of the PNS,
i.e., $\mpns/\rpns^2$.
As for the $f$ and $p$ mode,
the frequencies depend on the square root of the mean density
inside the shock.
Moreover,
\cite{Sotani24} indicates that the average density
is more suitable for universally expressing the CCSN
$g_1$- and $f$-mode GWs,
than the surface gravity of the PNSs.
\cite{Mori23} performed a long-term CCSN simulation
from core collapse to 20-seconds-later postbounce
and found that using compactness $\mpns/\rpns$ as a fitting variable
leads to a slightly better long-term fit
compared to the surface gravity and the mean density.
In addition,
\cite{Raduta20} confirmed
the $\bar{I}-\Lambda-\bar{Q}-C$ relations for PNSs
while keeping the entropy per baryon and lepton fraction constant.
\cite{Guedes24} demonstrated that the $\bar{I}-\Lambda-\bar{Q}$
and $\Omega_f-\Lambda$ relations are approximately valid for PNSs,
with deviations below approximately $10\%$
for postbounce time exceeding roughly $0.5\,$s.

After the detection of GWs from the inspiral phase of GW170817,
the study of BNSM simulations has flourished significantly.
During the merger, the densities and temperatures increase
and temperatures could reach several tens of MeV
in the postmerger phase \cite{Bauswein10b,Figura20,Figura21},
which will significantly influence the dynamical processes
\cite{Oechslin06,Baiotti08,Bauswein10a,Bauswein10b}.
The so-called $\Gamma$ law,
which incorporates ideal-gas thermal contributions into the cold EOS,
is widely used in BNSM simulations to account for finite-temperature effects
\cite{Figura20,Blacker20,Kochankovski22,Ilten22,Soultanis22,Blacker24}.
However, this simple approximation fails to consider
the impact of charge fraction across all densities.
Therefore,
it is vital to grasp the temperature dependence of the EOS more accurately.

Nevertheless,
numerical-relativity simulations have demonstrated that
the fundamental quadrupolar $f$ mode dominates
the GW spectrum of postmerger remnants
\cite{Shibata06,Hotokezaka13,Bernuzzi14,DePietri18}.
Some URs between
the peak frequency $f_\text{peak}$ during the postmerger phase
and the radius $R_{1.6}$ of a $1.6\ms$ NS
\cite{Bauswein12,Chatziioannou17,Torres-Rivas18},
the chirp mass $M_\text{chirp}$
\cite{Vretinaris20},
and the tidal deformability $\Lambda$
\cite{Rezzolla16}
have been suggested.
Specifically,
\cite{Bauswein18} demonstrated
that the concurrent detection of $\Lambda$ and $f_\text{peak}$
could offer a clear and definitive signature
of a strong first-order phase transition
occurring in the remnant during the merger process.
\cite{Blacker20} showed that a phase transition
breaks the UR between $M_\text{tot}, f_\text{peak}$,
and combined tidal deformability $\tilde{\Lambda}$
proposed by \cite{Bernuzzi15}.

The hydrodynamic simulation serves as a robust tool
for clarifying the internal mechanisms of forming compact objects,
while linear perturbation approaches provide
valuable insights into the physics
underlying the numerical results obtained from simulations.
However, obtaining the background structure of PNSs,
which is required for computing oscillation modes through linear analysis,
remains a significant challenge.
Notably, \cite{Ghosh23,Mu22,Kumar24}
studied linear $f$-mode oscillations within Cowling approximation
under conditions of constant entropy per baryon and lepton fractions
for nucleonic, hybrid, and hyperonic stars, respectively.
For the $f$-mode frequencies of cold NSs,
the Cowling approximation yields mass-dependent deviations
from the full solutions,
decreasing from $30\%$ to $10\%$ as mass increases \cite{Zheng25},
highlighting the necessity of investigating
$f$-mode oscillations within the frame of full general relativity (GR).
While \cite{Barman2025} recently demonstrated that
the $\Lambda-C$ and $\Omega_f-\Lambda$ URs
remain remarkably insensitive to nuclear saturation parameters
under constant entropy-per-baryon conditions for nucleonic NSs in full GR,
their model notably omitted neutrino trapping effects
-- a critical mechanism altering both thermal and structural evolution
in PNSs during early formation stages.
It is important to highlight that \cite{Burgio11}
conducted a detailed investigation of NRO modes,
specifically analyzing the $g_1$, $f$, and $p_1$ modes
under conditions of substantial entropy per baryon gradients
between the stellar core and envelope.

Simulations of CCSNe suggested that
the entropy per baryon in the PNSs remains
approximately constant (isentropic) throughout the cooling stage
\cite{Thompson03,Burrows13,Kumar24}.
In this work, we focus on this phase,
about several tens of seconds after the bounce,
extending until neutrino transparency is achieved.
Under these conditions,
we adopt the isentropic EOSs during the cooling stage
and systematically compare neutrino-free scenarios
with neutrino-trapping scenarios
to verify the URs applicable for cold NSs.
This analysis is performed
at constant lepton fractions of $Y_e=0.2,\,0.4,\ Y_\mu=0$,
where $Y_i \equiv x_i + x_{\nu_i}$ represents
the total lepton fraction for the species $i=e,\,\mu$.
Subsequently, to refine our predictions,
we incorporate more realistic thermal and neutrino-trapping profiles,
enabling a more accurate estimation of NRO properties.

In this work,
we extend our investigation to explore
the effects of temperature and neutrino trapping
on the NROs of PNSs,
building upon our previous studies of
cold NSs \cite{Sun21,Zheng23,Zheng25}
and PNSs \cite{Sun25}.
We employ the Brueckner-Hartree-Fock (BHF) theory for NM
\cite{Jeukenne76,Baldo99},
which incorporates realistic two-nucleon interactions
augmented by consistent microscopic three-nucleon forces
\cite{Grange89,Zuo02,Li08a,Li08b}.
This approach yields properties at nuclear saturation density
that align well with experimental data
\cite{Li08a,Li08b,Kohno13,Fukukawa15,Wei20,Burgio21}.
Furthermore,
the BHF framework satisfies observational constraints
derived from the analysis of the GW170817 event
and measurements by the NICER mission \cite{Wei20,Burgio21,Wei21}.
We adopt a recently developed finite-temperature BHF EOS \cite{Lu19}
and perform a comparative analysis
with a modern relativistic-mean-field (RMF) EOS \cite{Shen20},
which features a small symmetry energy slope.
In this framework,
we investigate the URs of PNSs in full GR.

This article is organized as follows.
In Sec.~\ref{s:eos} we review the formalism for the EOSs, i.e.,
the BHF theory and Shen RMF EOSs for PNSs.
In Sec.~\ref{s:osc} we introduce the macroscopic features and
the eigenvalue equations for the NROs of NSs.
Numerical results are given in Sec.~\ref{s:res},
and we draw the conclusions in Sec.~\ref{s:end}.
We use natural units $G=c=\hbar=k_\text{B}=1$ throughout the article.

\section{Equation of state for protoneutron stars}
\label{s:eos}

In the following we only provide a brief overview
of the formalism, referring to the relevant references
\cite{Burgio10,Lu19,Wei21,Liu22}.
The essential ingredient in the BHF many-body approach
is the in-medium reaction matrix $K$,
which is the solution of the Bethe-Goldstone equation
\be
 K(\rho,\xp;E) = V + \Re\sum_{1,2}V
 \frac{\ket{12} Q \bra{12}}{E-e_1-e_2+i0} K(\rho,\xp;E) \:
\label{e:k}
\ee
and
\be
 U_1(\rho,\xp) = \Re\sum_{2}
 n_2\expval{K(\rho,\xp;e_1+e_2)}{12}_a \:,
\label{eq:uk}
\ee
where $V$ is the nucleon-nucleon interaction potential,
$\xp \equiv \rho_p/\rho$ is the proton fraction,
and $\rho_p$ and $\rho$ are the proton and the total nucleon number densities,
respectively.
$E$ is the starting energy,
and
$e_i \equiv k_i^2\!/2m_i + U_i$ is the single-particle energy.
The multi-indices $1,2$ denote in general momentum, isospin, and spin.
In the case of finite temperature,
the Pauli operator $Q=(1-n_1)(1-n_2)$
and $n(k)$ is a Fermi-Dirac distribution.

Following the Bloch-DeDominicis approach
\cite{Bloch58,Bloch59,Baldo99,Baldo99b},
for a given density and temperature,
the Bethe-Goldstone equation can be solved self-consistently,
and the free energy density calculated exactly \cite{Burgio10,Liu22}.
A simplification of the scheme can be achieved
by disregarding the effects of finite temperature
on the single-particle potential,
which is the so-called frozen-correlations approximation
\cite{Burgio10,Lu19}.
Within this approximation,
the nucleonic free energy density can be simply expressed as
\be
 f = \sum_i \left[ 2\sum_k n_i(k)\left(\frac{k^2}{2 m_i}
 +\frac12 U_i(k) \right) -T s_i \right] \:,
\ee
where
\be
 s_i = -2\sum_k \left\{n_i(k) \ln n_i(k)
 + \left[1-n_i(k)\right] \ln\left[1-n_i(k)\right] \right\}
\ee
is the entropy density for the nucleonic species $i$.
From the nucleonic free energy density $f$
as function of temperature and particle number densities $\rho_i$,
one can compute all relevant observables in a thermodynamically consistent way,
such as chemical potentials $\mu_i$, pressure $p$, energy density $\eps$,
and entropy density $s$,
\bal
 \mu_i &= \frac{\partial f}{\partial \rho_i} \:,
\\
 p &= \rho^2 \frac{\partial(f/\rho)}{\partial\rho}
 = \sum_i \mu_i \rho_i - f \:,
\\
 s &= - \frac{\partial f}{\partial T} \:,~\eps = f + Ts \:,
\eal
where the subscript $i=n,p$ and $\rho=\rho_n+\rho_p$.

In the BHF approach,
the nucleon-nucleon interaction potential $V$ is the only necessary input.
In this work, we employ the Argonne $V_{18}$ (V18) \cite{Wiringa95}
potential,
supplemented with compatible microscopic three-body forces
\cite{Grange89,Zuo02,Li08a,Li08b}.
With this common prescription,
the saturation point of symmetric NM
and related properties can be reproduced correctly.
We use the convenient empirical parametrizations
for the free energy density given in \cite{Lu19,Wei21}.

The nonrelativistic BHF theory predicts a relatively stiff EOS
characterized by a large sound speed,
which can even become superluminal at extremely high densities.
To address this limitation,
we alternatively employ a RMF EOS,
which guarantees that the sound speed remains within causality constraints.
Specifically, we adopt the Shen 2020 EOS \cite{Shen20},
widely used in PNS modeling
and supernova simulations \cite{Shen02,Shen11,Shen19}.
This EOS employs the TM1e parametrization \cite{Bao14},
which modifies the density dependence of the symmetry energy
through the inclusion of additional $\omega-\rho$ coupling terms.
For this study, we use the EOS table published
in the CompOSE database \cite{compose}.

Compared to cold NSs,
PNSs exhibit a substantial neutrino abundance
owing to the neutrino trapping effect,
while the photon contribution must also be considered.
Under these conditions,
the hot and dense NM in the PNS is
constrained by beta equilibrium, charge neutrality, and fixed lepton fractions,
\bal
 & \mu_p + \mu_e = \mu_n + \mu_{\nu_e} \:,
\\
 & \mu_p + \mu_\mu = \mu_n + \mu_{\nu_\mu} \:,
\\
 & \rho_p - \rho_e - \rho_{\mu} =0 \:,
\\
 & Y_e = (\rho_e + \rho_{\nu_e})/\rho \:,
\\
 & Y_\mu = (\rho_\mu + \rho_{\nu_\mu})/\rho \:,
\eal
where $\rho_i$ and $\rho_{\nu_i}$ are the net electron/muon
and corresponding neutrino number densities, respectively.
Lepton fractions $Y_e=0.2,\,0.4$ and $Y_\mu=0$
are assumed for trapped matter.
In absence of neutrino trapping,
$\mu_{\nu_e}=\mu_{\nu_\mu}=0$,
and the lepton fractions $Y_i$ are not constrained.

The BHF approach provides only the EOS for the bulk-matter core region
without cluster formation,
and therefore has to be combined with a low-density crust EOS.
In contrast, the RMF EOS encompasses the description of non-homogeneous NM
within the PNS crust through the Thomas-Fermi approximation.
This framework posits that below a specific density threshold,
heavy nuclei can coexist with a free nucleon gas,
thereby reducing the free energy of the system.
The most stable configuration of NM is subsequently determined
by minimizing this free energy.
Consequently,
we employ the Shen EOS \cite{Shen20} for the low-density crust region,
independent of the core EOS.
To ensure a smooth transition between the two regimes,
we implement the log-space linear interpolation
across the density range of $0.05-0.08\,\fm3$.
To address neutrino trapping at low density,
we implement the following ``natural cutoff" procedure \cite{Burgio11}:
while maintaining the constant $Y_e$ and $Y_\mu$ across all densities,
the electron and muon neutrinos naturally vanish
at corresponding threshold densities.
For lower densities, the corresponding trapping condition is omitted.

\section{Neutron Stars}
\label{s:osc}

\subsection{Macroscopic features}

The Schwarzschild metric for a spherically-symmetric system is given by
\be
 ds^2 = - e^{\nu(r)}dt^2 + e^{\la(r)}dr^2
 + r^2(d\theta^2+\sin^2\!\theta d\phi^2) \:,
\label{e:ds2}
\ee
where $e^{\nu(r)}$ and $e^{\la(r)}$ are metric functions.
Given a specified EOS $p(\eps)$,
the mass-radius relation,
which represents one of the most fundamental
and essential macroscopic properties of compact stars,
can be derived by solving the Tolman-Oppenheimer-Volkoff (TOV) equations
\cite{Oppenheimer39,Tolman39}.
These equations describe the hydrostatic equilibrium of stars
within the framework of GR,
\bal
 \frac{dp}{dr} &= -\frac{e^\la}{r^2} (p+\eps)(m+4\pi r^3p) \:,
\label{e:dpdr}
\\
 \frac{dm}{dr} &= 4\pi r^2\eps \:,
\label{e:dmdr}
\\
 \frac{dm_B}{dr} &= 4\pi r^2 \rho m_n e^{\la/2}\:,
\eal
where $m$ is the enclosed gravitational mass,
$m_B$ is the enclosed baryonic mass,
and $m_n$ is the neutron mass.
One can obtain radius $R$ and gravitational mass $M=m(R)$
of a NS for a given central pressure or density
by solving these equations
with the boundary condition $p(R)=0$.
The corresponding metric functions are
\bal
 e^{\la} &= \frac{1}{1-2m/r} \:,
\\
 \frac{d\nu}{dr} &= -\frac{2}{p+\eps} \frac{dp}{dr} \:.
\label{e:metric}
\eal

GWs from a NS-NS or NS-BH merger
now provide valuable information on the EOS and internal structure of NSs,
in particular their dimensionless tidal deformability,
defined by the tidal Love number $k_2$ \cite{Hinderer08,Hinderer10},
\be
 \Lambda = \frac23 \frac{k_2}{\beta^5} = \frac{16}{15} \frac{z}{F}
\label{e:tdef}
\ee
with $\beta \equiv M/R$ the compactness of the star and
\bal
 z \equiv &\; (1-2\beta)^2 [2-y_R+2\beta(y_R-1)] \:,
\non\\
 F \equiv &\; 6\beta(2-y_R) + 6\beta^2(5y_R-8) + 4\beta^3(13-11y_R)
\non\\
          & + 4\beta^4(3y_R-2) + 8\beta^5(1+y_R) + 3z\ln(1-2\beta) \:.
\eal
$y_R=y(R)$ can be obtained by solving the additional differential equation
for $y(r)$ \cite{Lattimer06}
\be
 \frac{dy}{dr} = -\frac{y^2}{r} - \frac{y-6}{r-2 m} - r Q \:,
\ee
where
\be
 Q = 4\pi \frac{(5-y)\eps + (9+y)p + (p+\eps)(dp/d\eps)^{-1}}{1-2m/r}
 - \bigg(\frac{d\nu}{dr}\bigg)^2
\ee
with the boundary condition $y(r=0)=2$.

Moreover, in GR for a slowly uniformly rotating star,
the moment of inertia (MOI) can be expressed as \cite{Wei19}
\be
 I = \frac{w_R R^3}{6+2w_R} \:,
\label{e:moi}
\ee
where $w_R=w(R)$ can be obtained by solving the differential equation
\be
 \frac{dw}{dr} = 4\pi r \frac{(p+\eps)(4+w)}{1-2m/r} - \frac{w}{r}(3+w) \:,
\ee
with the boundary condition $w(0)=0$.
In this paper, the normalized dimensionless MOI is defined as
$\bar{I} \equiv I/M^3$.

\subsection{Nonradial oscillations and GW strain}

Thorne et al.\ developed a complete theory for NROs of NSs in GR
\cite{Thorne67}.
In this work,
we study only even-parity perturbations of the Regge-Wheeler metric,
\bal
 ds^2 =& -e^{\nu(r)} \big[
 1 + r^l H_0(r) e^{i\om t} Y^l_m(\theta,\phi) \big] dt^2
\non\\
       & +e^{\la(r)} \big[
 1 - r^l H_0(r) e^{i\om t} Y^l_m(\theta,\phi) \big] dr^2
\non\\
 & + \big[ 1 - r^l K(r) e^{i\om t} Y^l_m(\theta,\phi) \big]
 r^2 ( d\theta^2 + \sin^2\!\theta d\phi^2 )
\non\\
 & -2i\om r^{l+1} H_1(r) e^{i\om t} Y^l_m(\theta,\phi) dt dr \:,
\eal
where $Y^l_m(\theta,\phi)$ are the usual spherical harmonics,
$H_0$, $H_1$, and $K$ are metric perturbation functions, and
$\om = 2\pi f + i/\tau$ is the complex eigenfrequency of the NRO,
with $\tau$ the damping time of the GW.
The perturbation of the fluid in the star is described
by the Lagrangian displacement vector $\xi^\al$ in terms of
the dimensionless eigenfunctions $W(r)$ and $V(r)$
of the radial and transverse fluid perturbations,
\bal\label{e:xi}
 \xi^r      &= r^{l-1} e^{-\la/2} W Y^l_m \,e^{i\om t} \:,
\non\\
 \xi^\theta &= - r^{l-2} V \partial_\theta Y^l_m \,e^{i\om t} \:,
\non\\
 \xi^\phi   &= - r^{l-2} (\sin\theta)^{-2} V
 \partial_\phi Y^l_m \,e^{i\om t} \:.
\eal

In this work,
we employ the integration method developed by Lindblom and Detweiler
\cite{Lindblom83,Detweiler85}
to compute the nonradial oscillation modes.
Specifically,
the coupled system of linear equations
governing both the metric and fluid perturbations
is numerically integrated throughout the star interior,
subject to appropriate boundary conditions at the center and surface.
Further,
the perturbations of the metric outside the star
with the disappearance of fluid perturbations
can be described with a unique function $Z$,
which satisfies the Zerilli equation \cite{Zerilli70}.
This differential equation is numerically integrated
from the stellar surface outward to infinity,
with the physical constraint that the solution for $Z$
must represent a purely outgoing wave at infinity.
To determine the frequencies of quadrupole $(l=2)$ $f$-mode oscillations
we use the numerical methods developed in our previous work \cite{Zheng25}.

To determine the amplitude of the GW strain,
which depends on the oscillation amplitudes,
it is necessary to use the total energy of oscillation for normalization.
In Newtonian approximation,
the oscillation energy per radial distance
of an eigenmode is \cite{Thorne69b}
\be
 \frac{dE_\text{osc}}{dr} = \frac{\om^2}{2} (p+\eps) e^{(\la-\nu)/2}
 r^4 \left[ W^2 + l(l+1)V^2 \right] \:.
\label{e:dedr}
\ee
Generally,
the $f$-mode oscillation is considered to
be efficient at extracting energy at the level of
oscillations and driving the emission of GWs, thus
$E_\text{GW} \approx E_\text{osc}$.
By combining the oscillation energy
with the lowest-order post-Newtonian quadrupole formula
for the GW radiation power $P_\text{GW}$ \cite{Thorne69b,Reisenegger92},
one can derive an estimate for the GW damping time,
\be\label{e:taues}
 \taues \approx \frac{2E_\text{GW}}{P_\text{GW}} \:.
\ee
This approximation demonstrates remarkable accuracy,
yielding relative deviations within $7\%$
when compared to the full solutions for cold NSs \cite{Zheng25}.
The amplitude of the GW strain $h_+$ is also dependent on GW radiation power.
After some mathematical transformations and approximations,
it can be written as \cite{Zheng25}
\be
 h_+
 = \frac{\sqrt{15P_\text{GW}}}{\om D} \:
 = \frac{\sqrt{30E_\text{GW}/\taues}}{\om D} \:,
\label{e:hplus}
\ee
where $D$ is the distance to the source
and $\om$ is frequency of the NRO mode.
The detailed formulas and computational procedures
can be found in our previous work \cite{Zheng25}.
In this paper,
we use the damping time obtained from the full GR
to replace $\taues$ in Eq.~(\ref{e:hplus}).

\section{Numerical results}
\label{s:res}

\begin{figure}[t]
\vskip-5mm
\centerline{\includegraphics[width=0.5\textwidth]{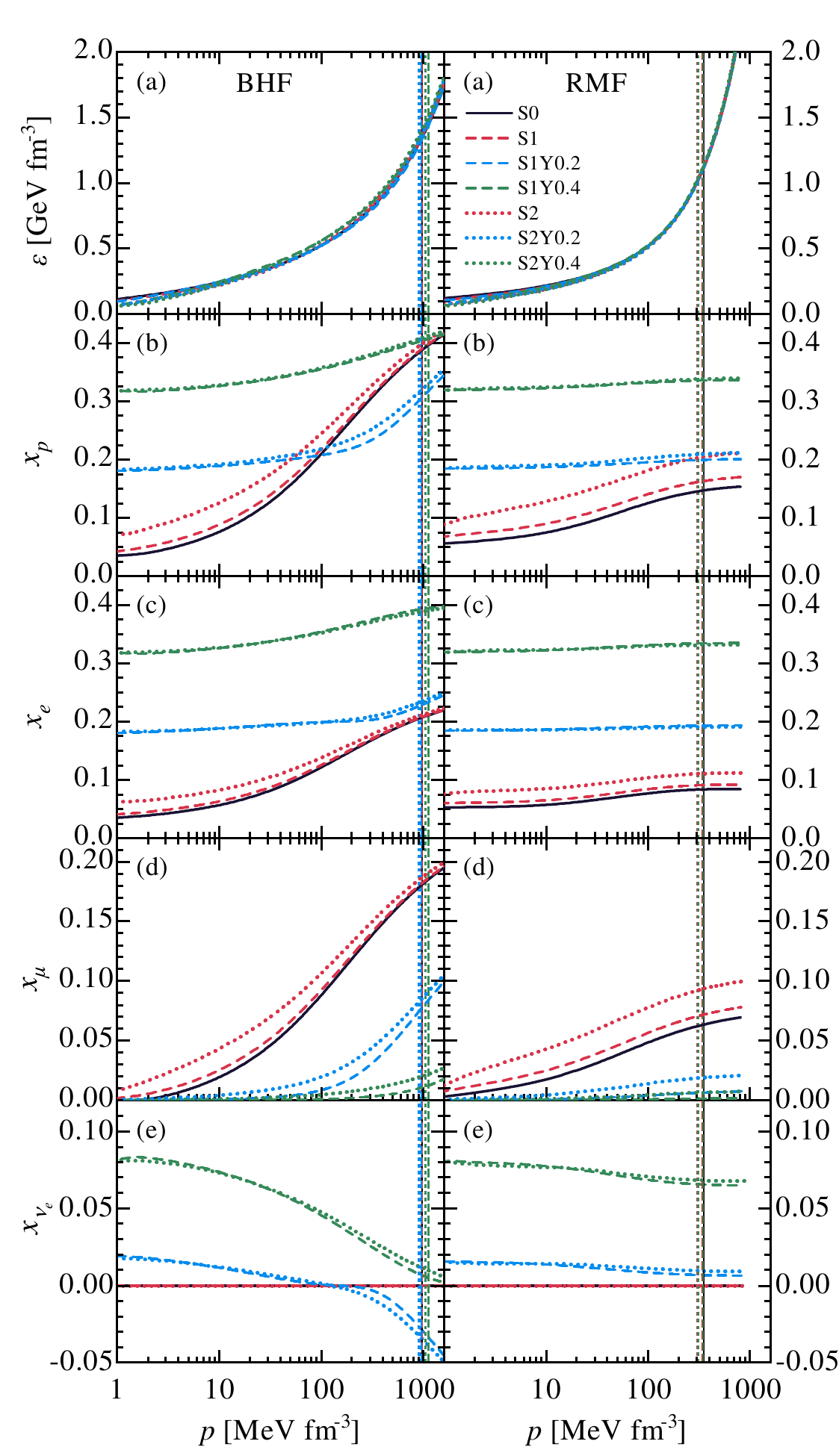}}
\vskip-3mm
\caption{
The energy density (a), proton fraction (b), electron fraction (c),
muon fraction (d),
and electron neutrino fraction (e)
of (P)NSs
as functions of pressure with BHF (left panels) and RMF (right panels) EOS,
for various values of $S/A$ and lepton fractions,
see the text for a detailed description of the notation.
The vertical lines indicate $\mmax$ configurations.
}
\label{f:eos}
\end{figure}

\begin{figure}[t]
\vskip-5mm
\centerline{\includegraphics[width=0.5\textwidth]{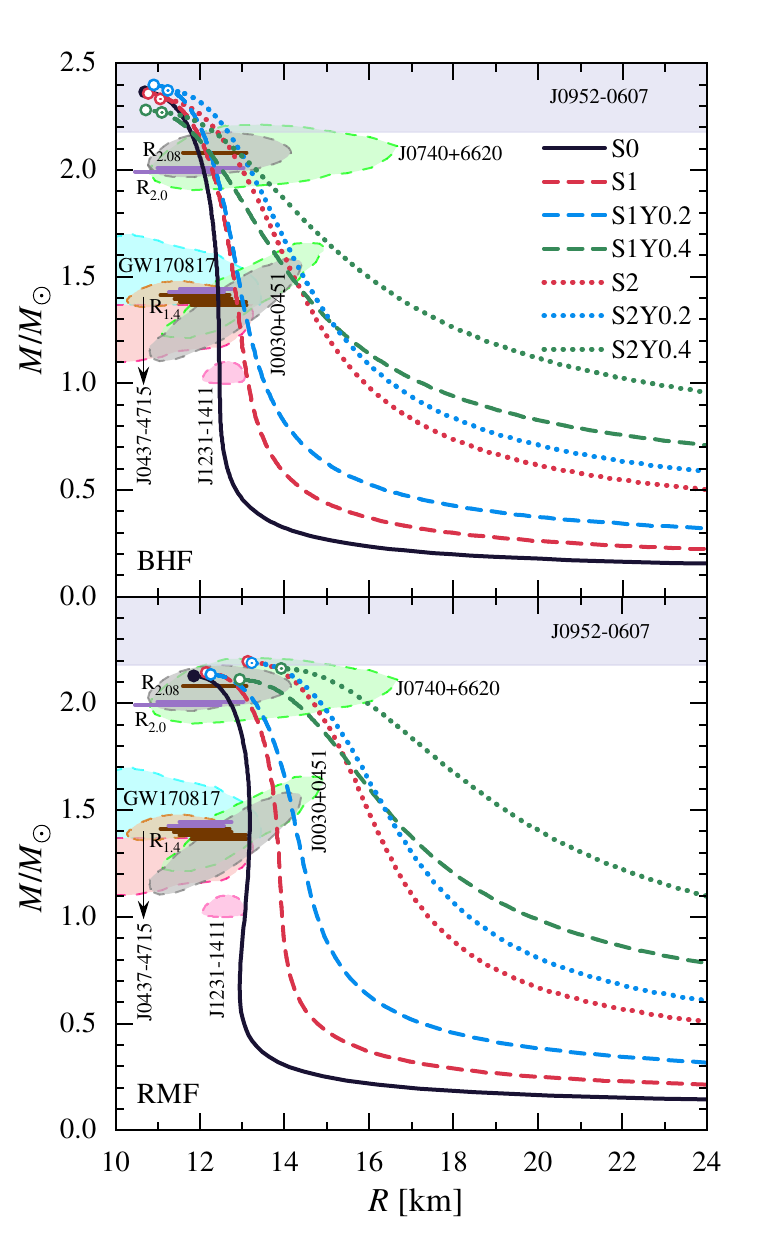}}
\vskip-5mm
\caption{
The gravitational mass--radius relations of (P)NSs
obtained with BHF (upper panel) and RMF (lower panel) EOSs.
The maximum-mass configurations are indicated by circle markers.
The horizontal bars indicate the limits on
$R_{2.08}$, $R_{2.0}$, and $R_{1.4}$
obtained in the combined NICER+GW170817 data analyses
of \cite{Miller21,Pang21,Raaijmakers21} (brown bars)
and the recent \cite{Rutherford24} (lilac bars).
The mass range of the heaviest currently known NS J0952-0607 is also shown.
The notation is as in Fig.~\ref{f:eos}.
}
\label{f:m-r}
\end{figure}

\subsection{Equation of state}

In this article,
we adopt the isentropic BHF EOS with V18 potential
and the Shen 2020 RMF EOS for NM.
The details of the EOSs are
comprehensively demonstrated in our previous work \cite{Sun25}. 
Here we present in Fig.~\ref{f:eos}
the energy density (a), proton fraction (b), electron fraction (c),
muon fraction (d),
and  electron neutrino fraction (e),
as functions of pressure.
Results for $S/A=0,1,2$
(denoted as ``S0, S1, S2")
and $Y_e=0.2,0.4$, $Y_\mu=0$ for neutrino trapping
(denoted as ``Y0.2,Y0.4")
are compared.
The different line styles represent distinct values of entropy per baryon,
while the color indicates various neutrino trapping conditions.
The static $\mmax$ configurations
are indicated for all cases by vertical lines.

The panels (a) show a stiffening of the EOS in the low-pressure regimes
with increasing either lepton fraction or entropy per baryon.
In the high-pressure regime,
for the RMF EOS,
increasing lepton fraction leads to slight softening,
whereas the BHF EOS exhibits stiffening at Y0.2 but softening at Y0.4
\cite{Sun25}.

These modifications due to temperature and in particular trapping
are the consequence of a delicate balance between two effects:
(i) a large imposed lepton fraction $Y_e$ leads in general to a reduction
of the proton fraction, i.e.,
nuclear matter becomes more symmetric and the nuclear pressure is reduced.
(ii) in compensation, the lepton ($e,\mu,\nu$) fractions are grossly modified:
$x_e$ strongly increases,
$x_\mu$ is reduced due to the $Y_\mu=0$ condition,
and there are additional $x_\nu$ contributions to the leptonic pressure.
The overall effect of all these changes is a relatively small modification
of the total pressure, which can be of both signs.

The above changes can be clearly seen in panels (b-e)
for the particle fractions:
In the neutrino-free case,
the proton fraction (b) of the RMF EOS at large pressure
remains much smaller than in the BHF case.
Imposing a fixed lepton fraction flattens
the $x_p$-pressure trends for both EOSs,
and generally increases $x_p$,
apart from a prominent crossover
between neutrino-free and Y0.2 cases of the BHF EOS.
This forced increase of nuclear asymmetry due to the trapping condition
underlies the non-monotonicity observed in the BHF EOS.
The electron fraction $x_e$ (c) increases substantially,
the muon fraction $x_\mu$ (d) is reduced,
and there is a substantial neutrino fraction (e).

Thus, the two EOSs considered feature very different intrinsic properties,
mainly related to their different proton fractions,
and it will be interesting to compare their predictions
regarding stellar oscillations.

\subsection{Macroscopic features of protoneutron stars}

In Fig.~\ref{f:m-r} we show the gravitational mass--radius relations of (P)NSs
with the same legend as in Fig.~\ref{f:eos}.
The mass-radius results of the NICER mission for the pulsars
J0030+0451 \cite{Riley19,Miller19,Vinciguerra24},
J0740+6620 \cite{Riley21,Miller21,Pang21,Raaijmakers21,Salmi24,Dittmann24},
PSR J0437-4715 \cite{Choudhury24},
and PSR J1231-1411 \cite{Salmi24b}
are also plotted in the figure.
The combined analysis of those pulsars together with GW170817
yields improved limits on
$R_{2.08}$ \cite{Miller21},
$R_{2.0}$ \cite{Rutherford24}, and
$R_{1.4}$ \cite{Pang21,Raaijmakers21,Rutherford24},
which are shown as horizontal bars.

The influence of entropy per baryon
and lepton fraction on the gravitational mass-radius relation of NSs is
clearly demonstrated in the figure.
Higher entropy states result in systematically larger radii
compared to the cold case for a given gravitational mass,
reflecting the thermal pressure contribution to the EOS.
Similarly, an increase in lepton fraction results in larger radii
at a fixed mass,
attributed to the stiffening of the EOS in the lower-pressure region,
as depicted in Fig.~\ref{f:eos}(a).
These thermodynamic effects are particularly pronounced in the lower mass range.
For a detailed discussion, see \cite{Burgio10,Wei21,Liu22}.

\begin{figure}[t]
\vskip-1mm
\centerline{\includegraphics[width=0.5\textwidth]{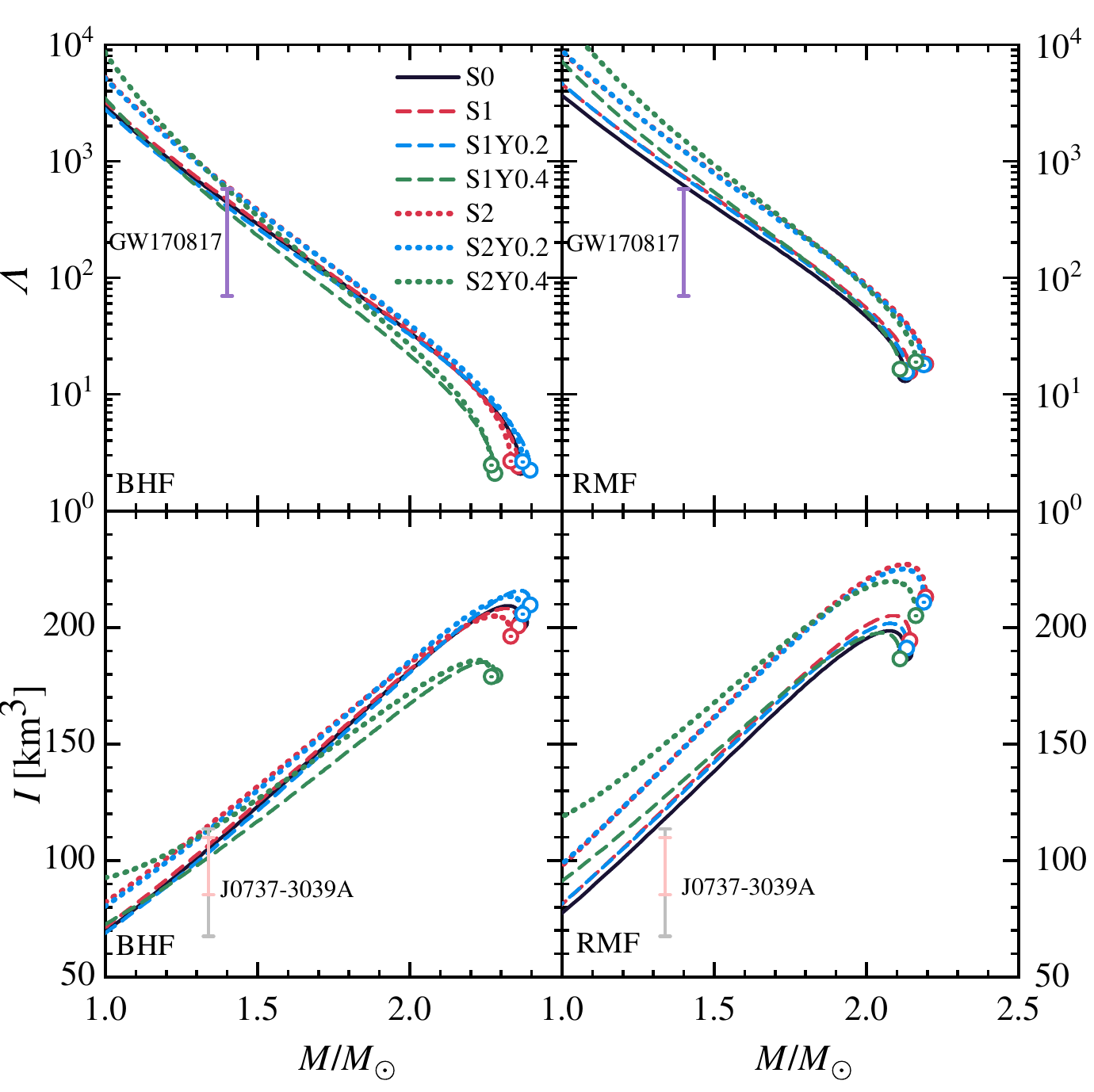}}
\vskip-4mm
\caption{
The dimensionless tidal deformability (upper panels)
and the MOI (lower panels) of (P)NSs vs gravitational mass $M$
obtained with BHF (left panels) and RMF (right panels) EOSs.
Observational values for GW170817 and the pulsar J0737-3039A are shown,
see text.
The notation is as in Fig.~\ref{f:eos}.
}
\label{f:tdmoi-m}
\end{figure}

\begin{figure}[t]
\vskip-1mm
\centerline{\includegraphics[width=0.5\textwidth]{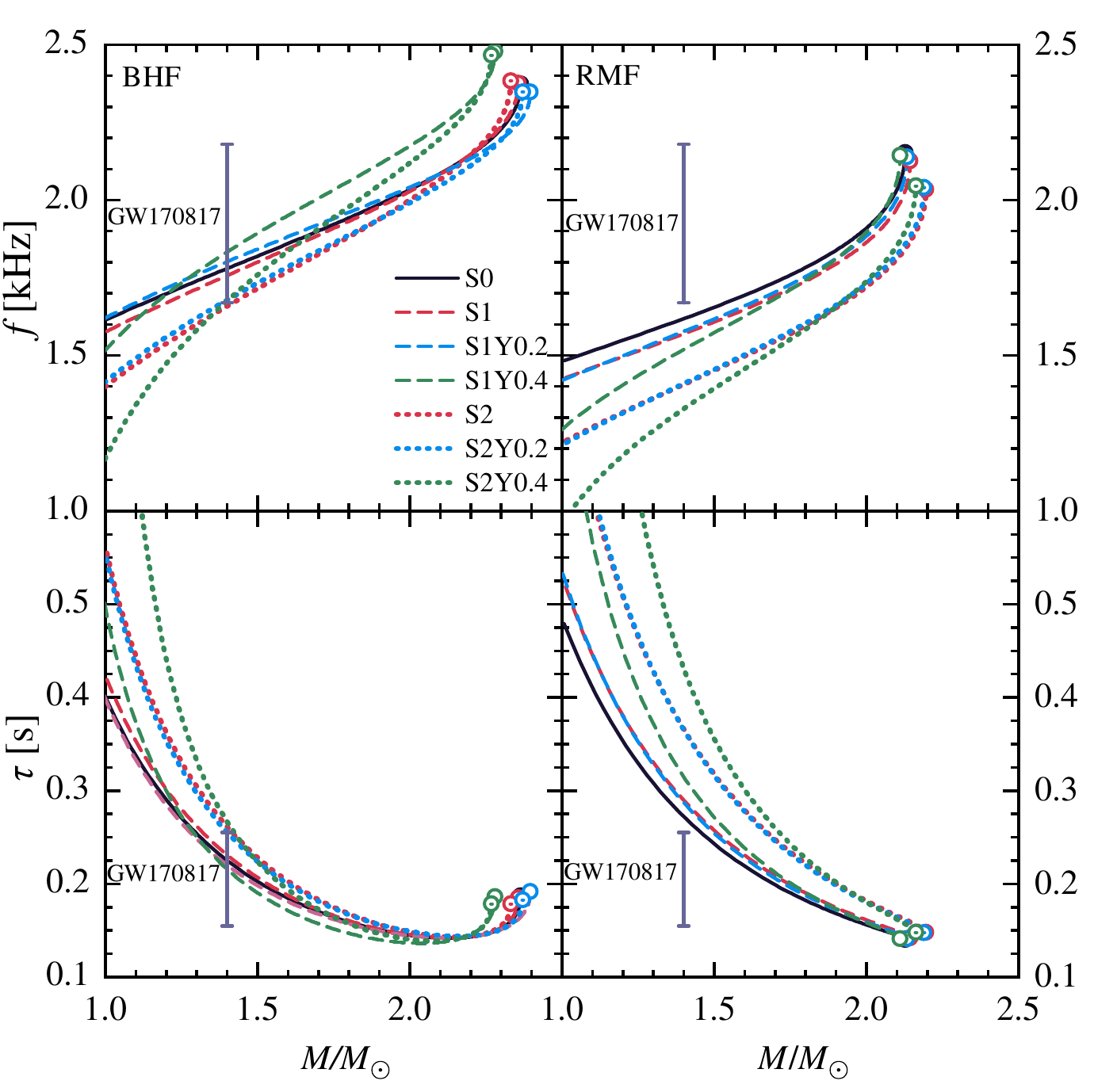}}
\vskip-4mm
\caption{
The GW frequencies (upper panels) and damping times (lower panels)
of $(l=2)\ f$-mode oscillations vs gravitational mass $M$
obtained with BHF (left panels) and RMF (right panels) EOSs.
The vertical bars indicate the GW170817 constraint \cite{Wen19}.
}
\label{f:ftau-m}
\end{figure}

\begin{figure}[t]
\vskip-4mm
\centerline{\includegraphics[width=0.5\textwidth]{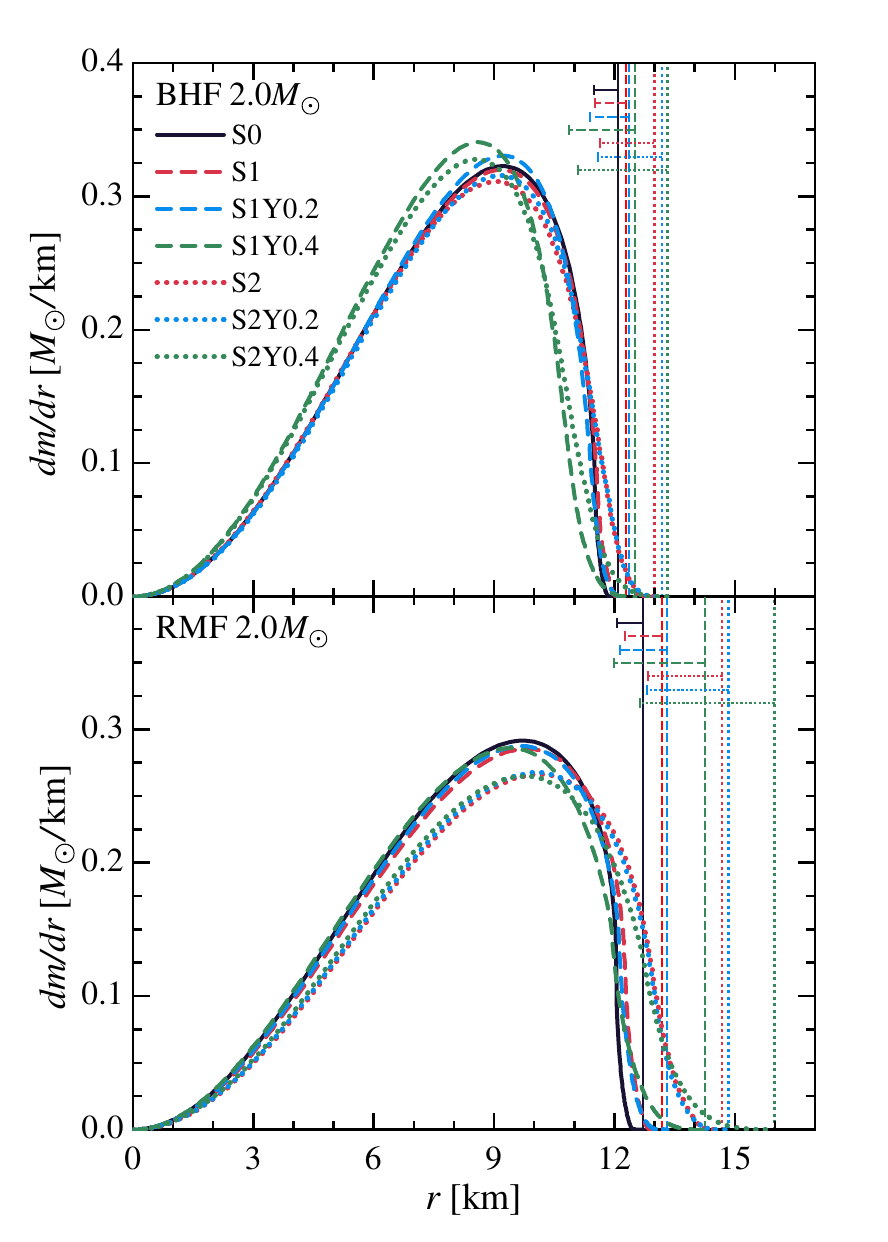}}
\vskip-4mm
\caption{
The radial mass distribution
for $2\ms$ (P)NSs
obtained with BHF (upper panels) and RMF (lower panels) EOSs.
The vertical lines indicate the radius $R$
and the horizontal bars the extension of the crust.
}
\label{f:dmdr}
\end{figure}

\begin{figure}[t]
\vskip-4mm
\centerline{\includegraphics[width=0.5\textwidth]{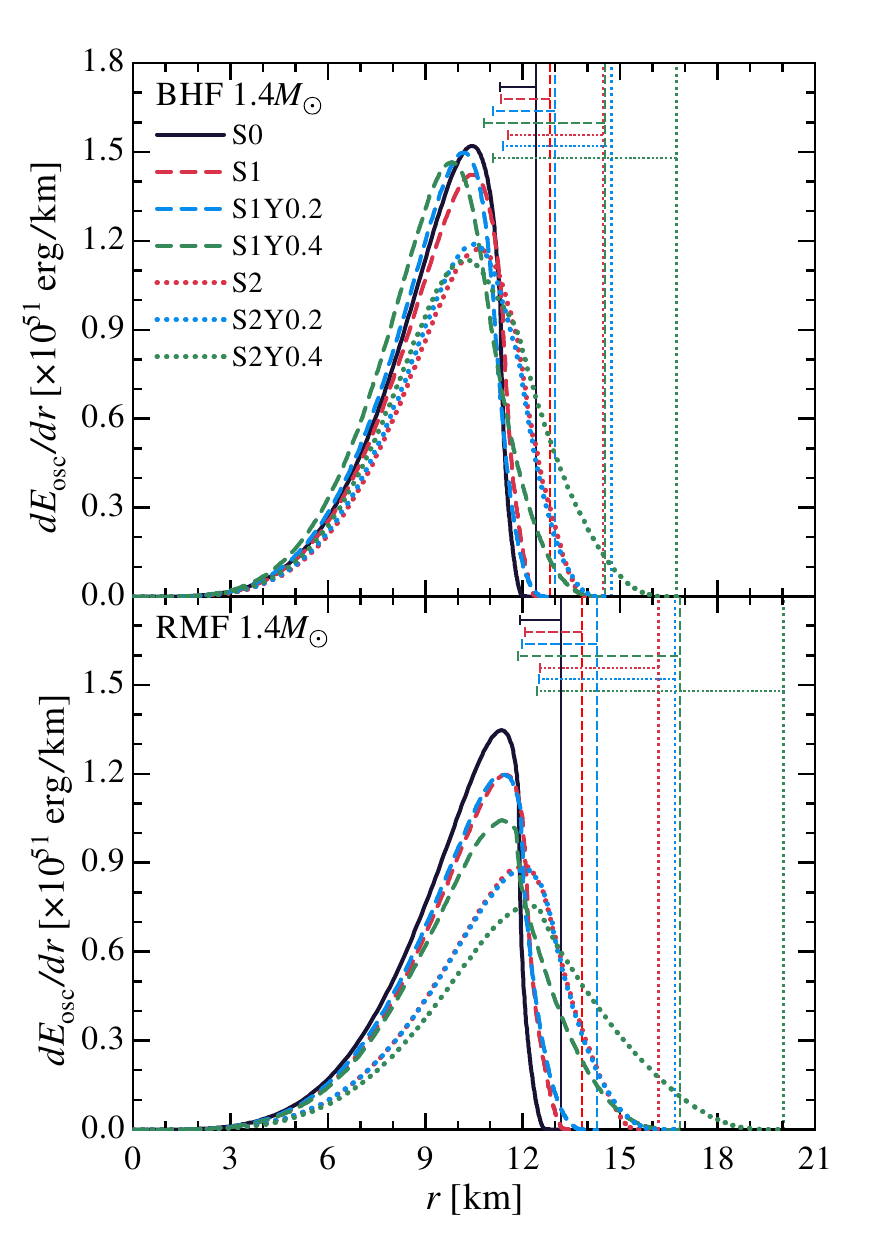}}
\vskip-4mm
\caption{
Same as Fig.~\ref{f:dmdr}, for
the radial distribution of oscillation energy, Eq.~(\ref{e:dedr}),
for $1.4\ms$ (P)NSs.
}
\label{f:dedr}
\end{figure}

\begin{figure}[t]
\vskip-5mm
\centerline{\includegraphics[width=0.52\textwidth]{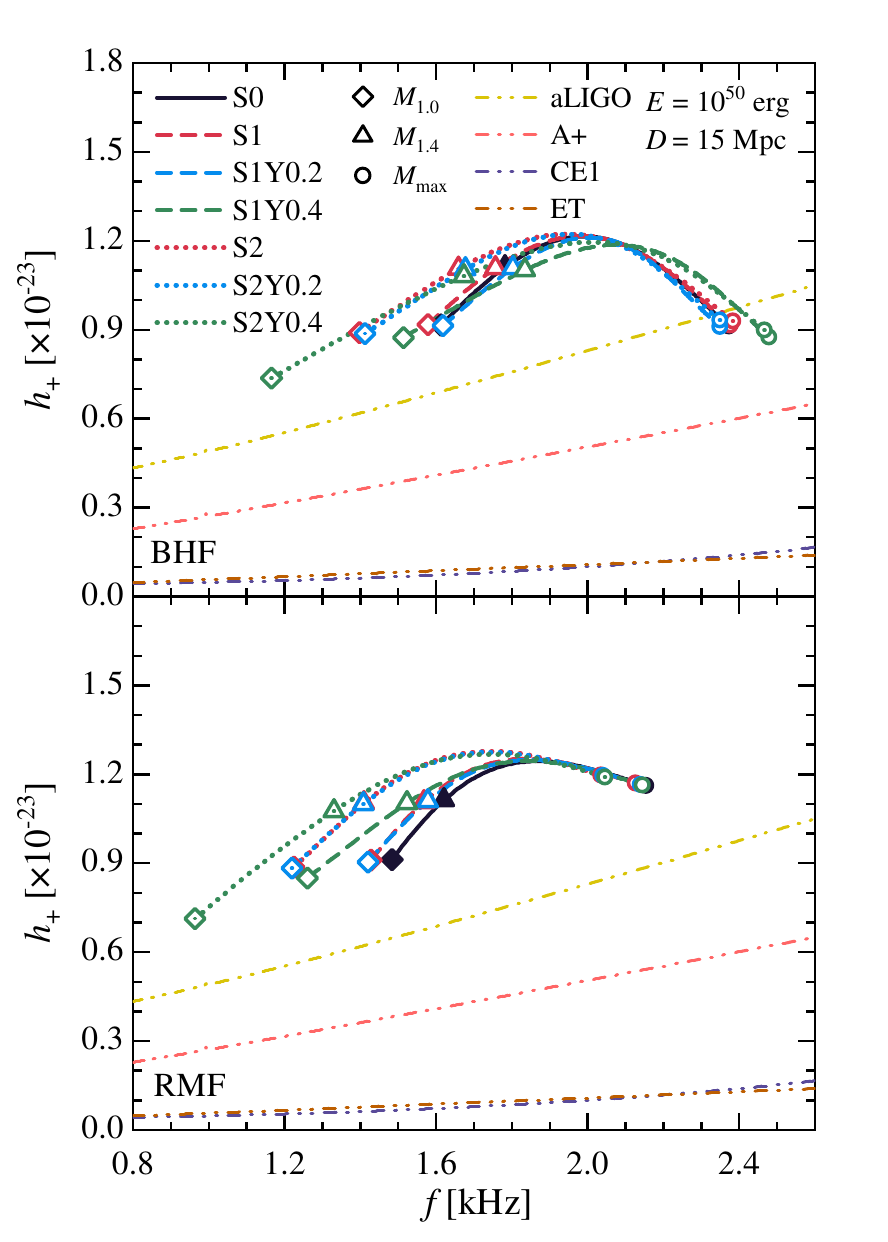}}
\vskip-4mm
\caption{
The peak mplitude of the GW strain $h_+$,
Eq.~(\ref{e:hplus}),
for a total GW energy $E=10^{50}\;$erg
and distance $D=15\;$Mpc,
emitted at frequency $f$ from a $f$-mode oscillation.
Stellar configurations of fixed $M=1\ms,1.4\ms,\mmax$ are indicated by markers.
The double-dotted lines are sensitivities of
Advanced LIGO (aLIGO), A+, Cosmic Explorer (CE1),
and Einstein Telescope (ET).
}
\label{f:h-f}
\end{figure}

In Fig.~\ref{f:tdmoi-m}
we show the dimensionless tidal deformability $\Lambda$, Eq.~(\ref{e:tdef}),
(upper panels),
and the MOI $I$, Eq.~(\ref{e:moi}),
(lower panels) as functions of the (P)NSs gravitational mass.
The constraint $\Lambda_{1.4}=190^{+390}_{-120}$
obtained by the analysis of the observational data
from GW170817 and its electromagnetic counterpart
\cite{Abbott18} (purple bars),
as well as the constraints on the MOI
of the pulsar J0737-3039A
based on the GW170817 tidal measurement
\cite{Landry18} (grey bars)
or by using the multi-messenger constraints on the radius \cite{Dietrich20}
in combination with the radius-MOI relation \cite{Lattimer19} (pink bars)
are also shown in the figure.
Tidal deformability in (P)NSs quantifies
their degree of distortion under external tidal fields.
The figure reveals that while variations
in entropy per baryon and lepton fraction
induce moderate changes of the tidal deformability,
they do not substantially modify the fundamental
inverse correlation between tidal deformability and stellar mass.
Overall,
the influence of entropy and lepton fractions is more prominent in
the results obtained from the RMF EOS compared to BHF EOS.
An analogous behavior is observed in the moment of inertia (lower panels),
demonstrating consistent thermal dependence
across different physical properties of PNSs.

In particular we note that thermal effects and trapping
always increase both $\Lambda$ and $I$ for the RMF EOS,
which is consistent with the increased radii in Fig.~\ref{f:m-r}.
However, with BHF EOS the Y0.4-trapped stars
have smaller $\Lambda$ and $I$ over a large mass range,
although their radii are also increased.
The reason is seen in the radial mass distribution
$dm/dr=4\pi\eps$ shown in Fig.~\ref{f:dmdr}:
For the BHF EOS the centroid of the trapped mass distributions
is at significantly lower radius than for the nontrapped stars
(although the low-density crust has a larger extension with trapping),
i.e., those stars are more compact,
which is not the case for the RMF EOS.
This feature is related to the change of the proton fraction
within the stellar interior
induced by the fixed lepton fraction constraint:
The proton fraction of betastable matter is above 0.3 at large density
for the BHF EOS \cite{Lu19},
whereas it remains around 0.1 for the RMF \cite{Shen20},
see Fig.~\ref{f:eos}.
Therefore the trapping condition Y0.4 causes a much larger softening for
the RMF due to the matter becoming more symmetric.

\subsection{$f$-mode oscillations}

We now turn to a comparison of the $f$-mode frequencies
for different conditions.
In Fig.~\ref{f:ftau-m},
we show the GW frequencies (upper panels) and
the damping times (lower panels)
of $(l=2)~f$-mode oscillations vs gravitational mass $M$
obtained with BHF (left panels) and RMF (right panels) EOSs.
We also show the constraints (vertical bars) for a canonical NS
obtained by using the deduced GW170817 tidal deformability
\cite{Wen19,Pratten20}.
Temperature and trapping effects are in line with those observed
in Figs.~\ref{f:m-r} and \ref{f:tdmoi-m}, namely,
less compact stars feature lower frequency and larger damping time.
This consistency highlights the strong correlation
between the characteristics of $f$ modes and macroscopic features
across different thermal conditions,
as observed in the study of cold NSs
\cite{Andersson98,Kokkotas99}.

In the same figure, one can see that the damping times in full GR
are of the order of a few tenths of a second.
In the majority of cases
(but not for the trapped BHF EOS at most masses),
PNSs exhibit longer damping times
compared to their cold counterparts,
indicating reduced GW radiated power
and consequently smaller GW strain amplitudes.
While this characteristic generally suggests greater observational challenges
for detecting $f$-mode GW signals from PNSs,
several compensating factors must be considered.
The substantially larger energy reservoir available in PNSs
for exciting oscillations,
combined with the enhanced probability of oscillation-triggering mechanisms
during the early evolutionary stages,
and the lower frequencies of GWs,
may effectively counteract the reduced GW radiation power.
Consequently,
despite the lower GW emission efficiency,
the overall detectability of PNS oscillations
through their $f$-mode GW signatures might not be substantially diminished.

To elucidate the characteristics
of $f$-mode oscillations within the star,
Figs.~\ref{f:dmdr} and \ref{f:dedr}
present respectively the radial distributions of
mass for $2\ms$ (P)NSs and
oscillation energy for $1.4\ms$ (P)NSs
[with normalization determined by the boundary condition $W(r=0)=1$
in Eq.~(\ref{e:xi})].
One can see that,
while temperature and in particular neutrino-trapping effects
substantially increase the stellar radius,
their impact on the core-crust interface radius remains small
(horizontal bars).
The maximum of mass and oscillation energy per distance
consistently occur in the outer core,
almost unaffected by thermal and neutrino trapping effects.
This suggests that $f$-mode oscillations
predominantly manifest in the stellar crust and outer core,
which potentially explains the existence of tighter URs
for $f$-mode oscillations,
as these relations primarily reflect the properties
of the crust and outer core rather than the uncertain inner core.

In order to accurately estimate GW strain amplitudes,
the determination of energy scales
associated with the emission process is essential.
In the case of a typical CCSN,
the total released energy is $\sim 10^{53}\;$erg,
while the kinetic energy of mass ejecta is $\sim 10^{51}\;$erg
\cite{Lugones21}.
We assume a conservative estimate
that approximately 10\% of the total available energy
is efficiently converted into $f$-mode GW radiation, i.e., $10^{50}\;$erg.
Choosing a typical distance $D\sim15\;$Mpc
(star in the Virgo cluster),
we show in Fig.~\ref{f:h-f} the GW strain amplitude
obtained with various EOSs
against the sensitivity curves of
Advanced LIGO (aLIGO) \cite{Abbott18b},
A+ \cite{A+},
Cosmic Explorer (CE1) \cite{CE&ET},
and Einstein Telescope (ET) \cite{Hild11}.

As evident from the figure,
the GW strain amplitude exhibits a non-monotonic dependence on
frequency, i.e., stellar mass,
displaying a slight decrease as the mass approaches the maximum mass.
Since $h_+ \sim 1/(f\sqrt\tau)$,
this behavior stems from the competing effects of
increasing oscillation frequency and decreasing damping time
with increasing mass,
see Fig.~\ref{f:ftau-m}.
The $f$-mode GW strain peak amplitudes for $1.4\ms$ (P)NSs
in the Virgo cluster
maintain a nearly constant value of approximately $1.1\times10^{-23}$
across different conditions,
falling within the detection capability of current GW detectors
(at least A+).

\begin{figure}[t]
\vskip-1mm
\centerline{\includegraphics[width=0.5\textwidth]{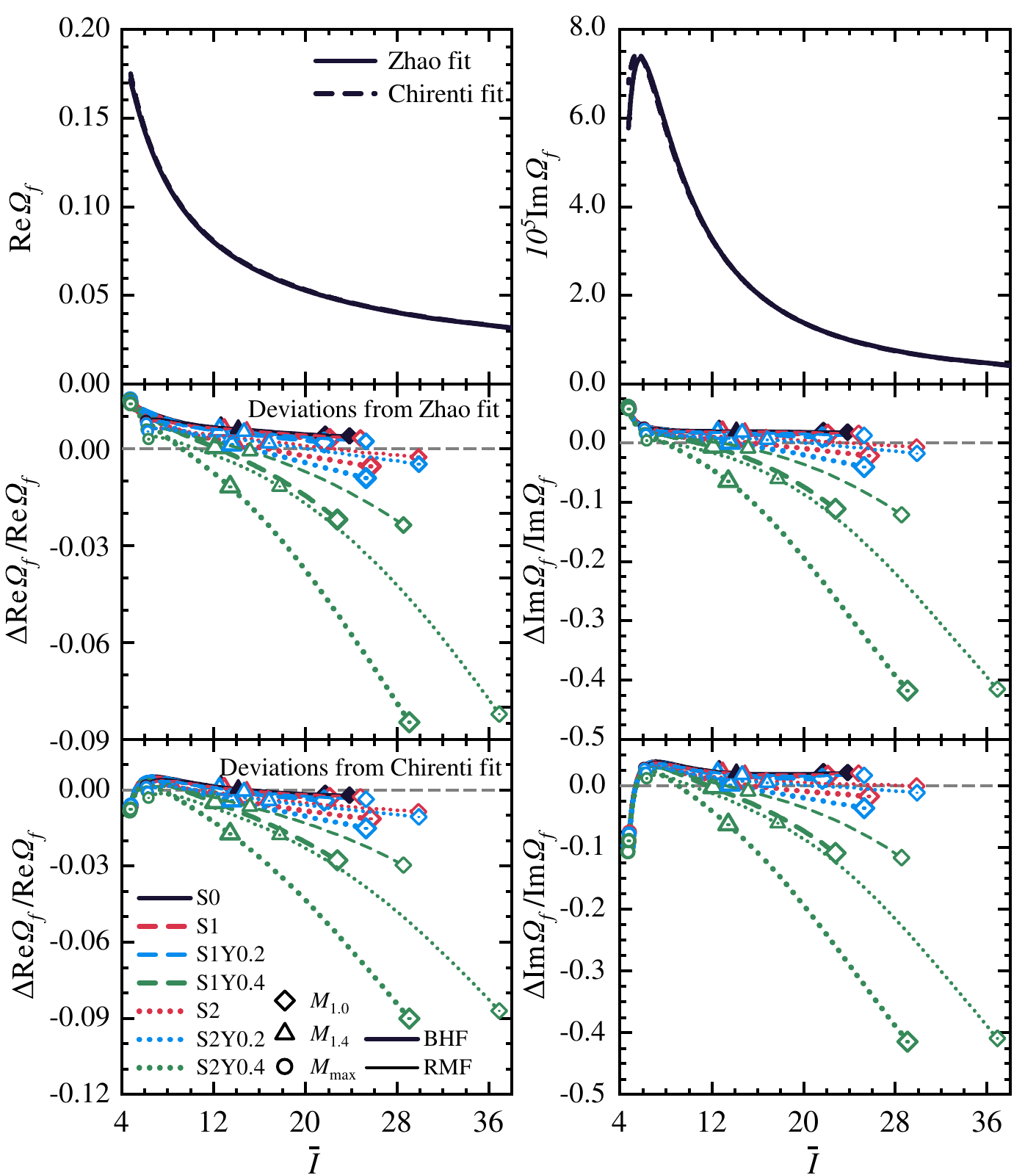}}
\vskip-2mm
\caption{
Top panels:
The URs of \cite{Zhao22b} (Zhao) and \cite{Chirenti15} (Chirenti)
between dimensionless $f$-mode frequency $\Omega_f$
(real and imaginary part)
and dimensionless MOI $\bar{I}$,
and their relative deviations from our considered EOSs
(lower panels).
Stellar configurations of fixed $M=1\ms,1.4\ms,\mmax$ are indicated by markers.
}
\label{f:mf-moi/m3}
\end{figure}

\begin{figure}[t]
\vskip-1mm
\centerline{\includegraphics[width=0.5\textwidth]{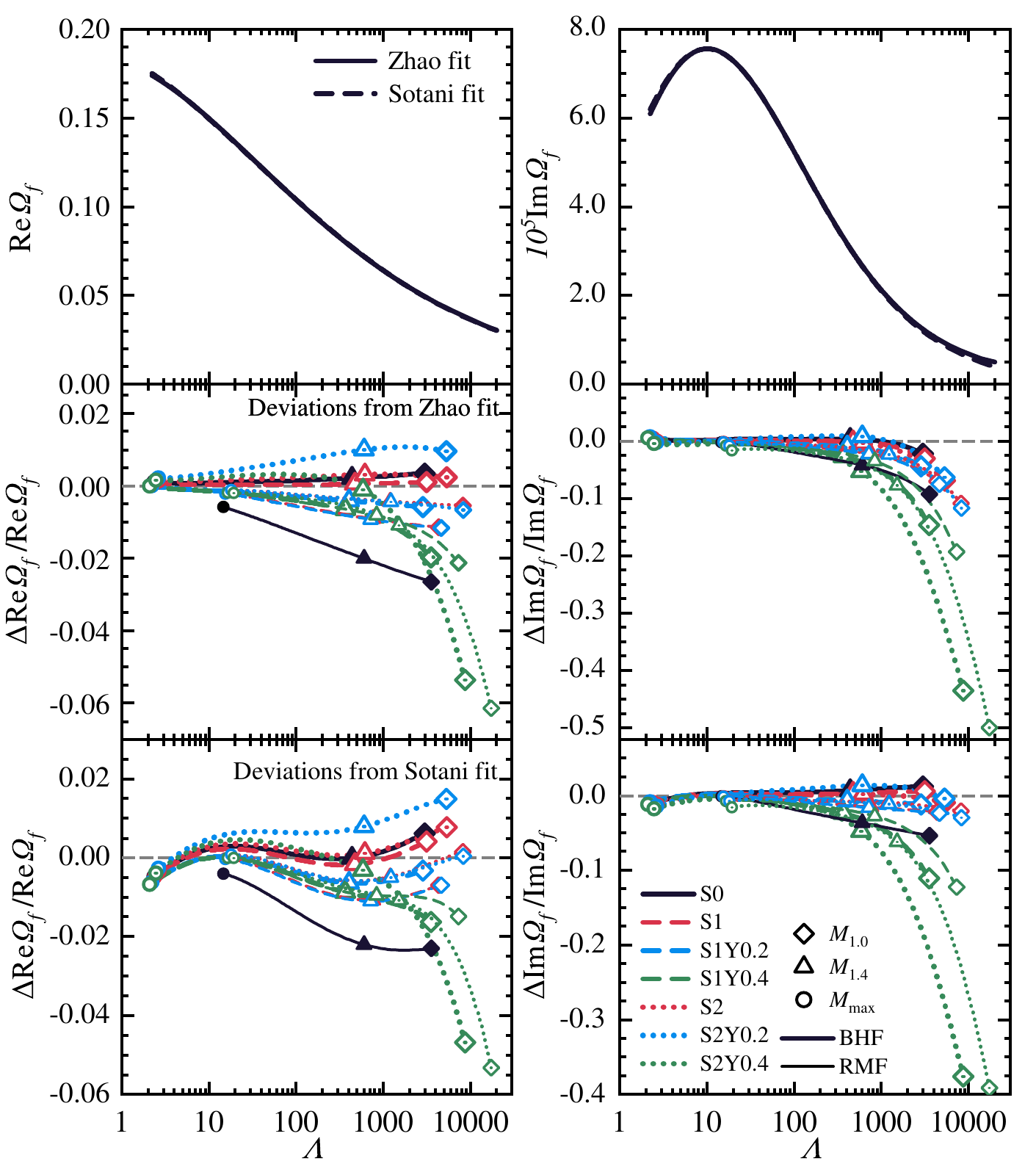}}
\vskip-2mm
\caption{
Same as Fig.~\ref{f:mf-moi/m3},
for the $\Omega_f(\Lambda)$ relation,
in comparison with URs of \cite{Zhao22b} (Zhao) and \cite{Sotani21} (Sotani).
}
\label{f:mf-lmbda}
\end{figure}

\subsection{Universal relations}

For cold NSs,
the URs connecting $f$-mode characteristics
to macroscopic properties of the star
have been thoroughly investigated and well established in the literature
\cite{Andersson98,Kokkotas99,Lau09,Chan14,Chirenti15,
Sotani21,Zhao22b,Pradhan22,Zheng25}.
Particularly noteworthy is the UR between the
dimensionless frequency $\Omega_f \equiv M\om_f$
and either dimensionless MOI $\bar{I} \equiv I/M^3$
\cite{Lau09,Chirenti15}
or dimensionless tidal deformability $\Lambda$
\cite{Chan14,Sotani21,Zhao22b}.
These relations,
which exhibit remarkable EOS independence,
demonstrate significant potential as robust tools
for extracting NS properties from $f$-mode GW observations.
However, for PNSs,
the validity and universality of these relations remain largely unexplored.
Notably, \cite{Barman2025} have recently demonstrated
that for nucleonic NSs with masses exceeding one solar mass,
the $\Omega_f(\Lambda)$ UR
maintains remarkable insensitivity to
both nuclear saturation parameters and entropy per baryon
under isentropic conditions in full GR.
Here, we examine whether these URs remain applicable
when accounting for both isentropic and neutrino trapping conditions
during the early postbounce phase,
thereby extending the analysis beyond the cold NS regime.

We will now check the compliance of our isentropic nucleonic EOSs
with these URs.
The results are displayed in Figs.~\ref{f:mf-moi/m3} and \ref{f:mf-lmbda}
for $\Omega_f(\bar{I})$ and $\Omega_f(\Lambda)$, respectively.
The upper panels show the URs for cold NSs given by
Zhao \cite{Zhao22b}, Chirenti \cite{Chirenti15}
and
Zhao \cite{Zhao22b}, Sotani \cite{Sotani21},
respectively,
whereas the middle and bottom panels show the relative deviations
$(\Omega_\text{EOS}-\Omega_\text{UR})/\Omega_\text{UR}$
of our various EOSs from the fits.
The figures display the range of URs applicable to cold NSs.
Notably, when accounting solely for entropy variations
(black and red curves),
the URs for the real (imaginary) parts of $\Omega_f$
are valid within about
1\% (2\%) for $\Omega_f(\bar{I})$
and
3\% (10\%) for $\Omega_f(\Lambda)$,
demonstrating the robustness of these relations
against moderate thermal effects.

However,
in the mass range below $1.4\ms$
the deviations from URs exhibit
a pronounced sensitivity to lepton fraction,
demonstrating an enhancement with increasing electron lepton fraction.
Most notably in the high-entropy S2Y0.4 case,
the deviations display a rapid mass-dependent amplification,
scaling inversely with stellar mass.
This behavior manifests coherently in both the macroscopic properties
(see Figs.~\ref{f:m-r}, \ref{f:tdmoi-m})
and the $f$-mode oscillation characteristics
(see Fig.~\ref{f:ftau-m}).
Therefore, in this low-mass
(high $\bar{I}$ and $\Lambda$)
range the URs become invalid due to the strong sensitivity
on the trapping condition,
as exposed in the previous figures.
Nevertheless,
for (P)NSs with masses exceeding $1.4\ms$,
the URs for the real (imaginary) parts of $\Omega_f$
are still valid within about
2\% (2\%) for $\Omega_f(\bar{I})$
and
3\% (8\%) for $\Omega_f(\Lambda)$.

Current theoretical understanding suggests that while future GW detectors
with enhanced sensitivity could precisely determine $f$-mode frequencies
from observed signals \cite{Kokkotas01c},
the damping times may remain more challenging
to measure with comparable accuracy.
CCSN simulations consistently identify $f$-mode oscillations
as the dominant frequency in PNS GW emissions,
with $g$~modes occasionally contributing
\cite{Morozova18,Sotani19,Sotani20,Wolfe23}.
Complementing these findings,
numerical-relativity studies have established that the quadrupolar $f$~mode
also dominates the GW spectrum of BNSM remnants
\cite{Shibata06,Hotokezaka13,Bernuzzi14,DePietri18}.
With the established URs and anticipated improvements
in constraints of the tidal deformability,
future observations will allow significantly more precise determinations
of both the $f$-mode frequency and damping time \cite{Wen19}.
Furthermore,
the combined analysis of tidal deformability measurements
from the premerger inspiral phase
and postmerger peak frequency $f_\text{peak}$,
which originates from quadrupolar $f$-mode oscillations,
provides a powerful tool for identifying quark-matter phases
in NS interiors \cite{Blacker20}.


\subsection{Stellar profiles}

\begin{figure}[t]
\vskip-5mm
\centerline{\includegraphics[width=0.45\textwidth]{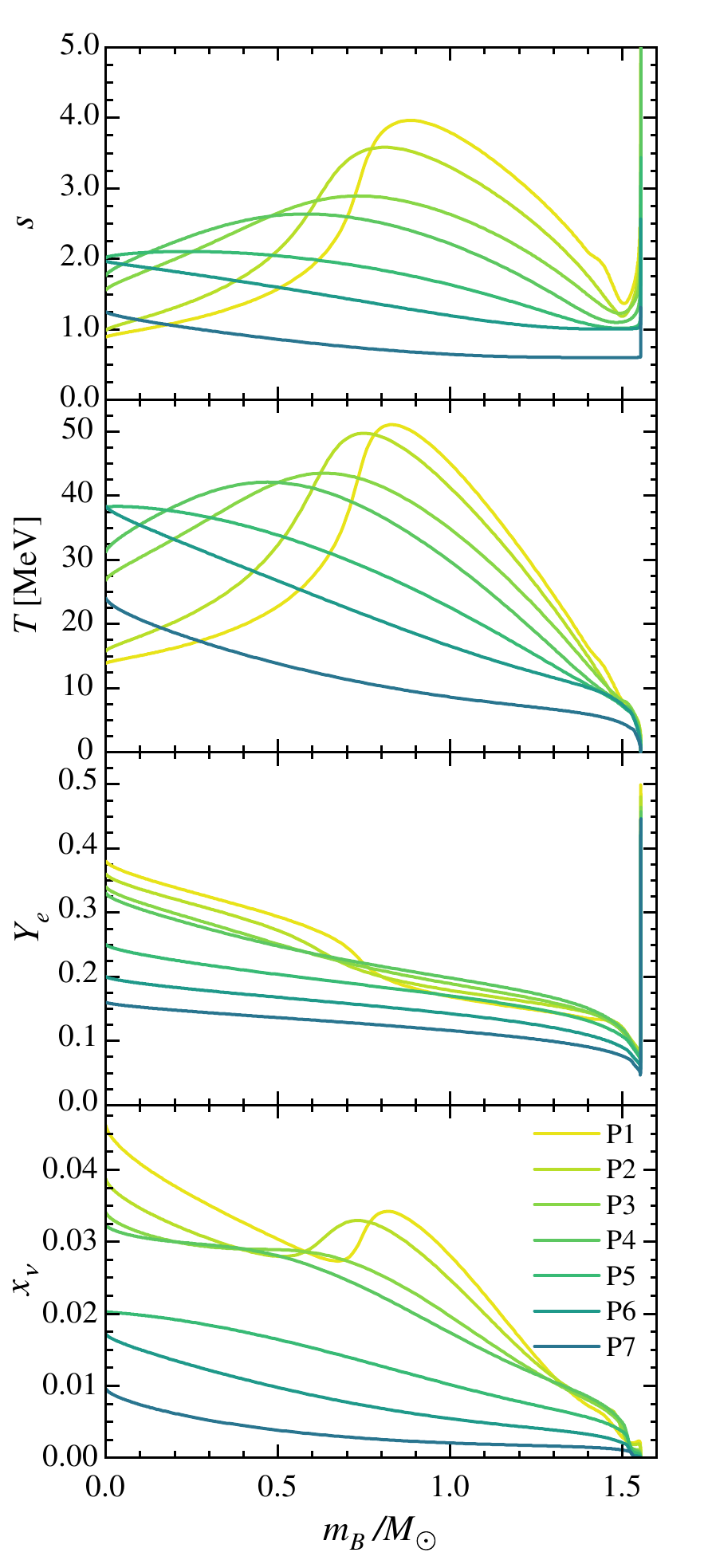}}
\vskip-5mm
\caption{
The entropy per baryon $s$,
temperature $T$,
lepton fraction $Y_e$, and
neutrino fraction $x_\nu$
as functions of the enclosed baryonic mass $m_B$
corresponding to a $M_B=1.55\ms$ star
with the BHF EOS,
for several time snapshots of Table~\ref{t:profile},
labeled by the different colors.
}
\label{f:frad}
\end{figure}

Recent three-dimensional simulations of CCSNe have revealed
that the accretion phase persists beyond $\sim 1\,$s after the bounce
\cite{Lentz15,Vartanyan18,Burrows19}.
Particularly noteworthy are the findings of \cite{Vartanyan18},
which demonstrate that while the majority of baryonic-mass accumulation
occurs within the initial $\sim 0.2\,$s,
both the baryonic mass and gravitational mass continue to increase significantly
even at timescales exceeding $1\,$s after core bounce.
This suggests that the PNS maintains a state of dynamic evolution
rather than hydrostatic equilibrium due to accretion.
To ensure the validity and stability of perturbation analysis,
which is fundamentally based on equilibrium stellar structure,
we adopt a conservative approach by selecting the epoch beginning at $2\,$s
after core bounce for our investigations.
This choice is motivated by the fact that by this evolutionary stage,
the PNS is more likely to be described within the quasi-static approximation,
with the timescale for significant structural changes
becoming substantially longer than
the characteristic oscillation periods of interest.

To achieve more accurate predictions for the $f$-mode characteristics,
it is essential to incorporate realistic thermal profiles
and neutrino trapping conditions during PNS evolution,
as demonstrated in recent studies
\cite{Pons99,Fischer10,Burgio11,Camelio17,Li21,Mori23,Sotani24}.
Current dynamical simulations of postbounce PNS evolution
reveal a characteristic structural evolution:
following core bounce, a low-entropy core emerges,
enveloped by a high-entropy mantle resulting
from shock-wave propagation through the outer PNS layers.
During the initial evolutionary phase,
neutrinos remain temporarily trapped within the core,
but subsequently diffuse outward into the envelope
over timescales of several seconds,
depositing substantial thermal energy throughout the star.
This neutrino-driven heating process elevates
core temperatures to several tens of MeV,
leading to an approximate doubling of the core entropy
while simultaneously reducing entropy in the envelope.
Over timescales of tens of seconds,
the PNS undergoes significant lepton number depletion,
accompanied by gradual cooling and homogenization of entropy gradients.
Concurrently, the average neutrino energy decreases
while their mean free path increases,
eventually reaching values comparable to the stellar radius
after approximately one minute.
This transition to neutrino transparency
marks the final evolutionary stage,
culminating in the formation of a cold NS.

In our previous work \cite{Burgio11,Sun25},
we modeled PNS configurations
with a low-entropy core and high-entropy envelope
smoothly connected via cubic interpolation,
assuming a constant lepton fraction throughout the star.
In this study,
we develop profiles of both entropy per baryon $s$
and electron lepton fraction $Y_e$
as functions of baryon number density $\rho$
to reproduce the stellar profiles presented in \cite{Roberts17},
\bal
 s(\rho) =&\; \smin + (\smax-\smin) \exp(-\frac{(\rho'-a)^2}{b\rho'})
 - h \log(\frac{\rho'}{\rho'+k}) \:,
\label{e:profile_s}
\\
 Y_e(\rho) =&\; Y_{s} + (Y_{c}-Y_{s}) \rho' \:,
\label{e:profile_y}
\eal
where
$\rho'\equiv\rho/\rho_c$ with
the central baryon number density $\rho_c$,
$\smin$ and $\smax$ are related to the minimum and maximum entropy per baryon
in the core, respectively,
and the electron lepton fractions at the stellar center and surface
are denoted as $Y_c$ and $Y_s$, respectively.
The parameters $a,b,h,k$
are used to manipulate the shape of the entropy function.

We present in Table~\ref{t:profile} the selected configurations
for a fixed baryonic mass $M_B=1.55\ms$,
corresponding to a cold NS with gravitational mass $M=1.4\ms$ using the BHF EOS.
The table lists both the eight parameters
and the estimated initial time $t$ of the profiles.
The selected entropy profiles and lepton fractions
are based on dynamical simulations from \cite{Pons99,Roberts17},
which employed various EOSs.
These EOS variations may introduce minor uncertainties
in the time evolution estimates presented in the final column.

\begin{table}[t]
\def\myc#1{\multicolumn{2}{c}{$#1$}}
\caption{
Stellar models of different entropy and lepton-fraction profiles
for a fixed baryonic mass $M_B=1.55\ms$,
corresponding to a gravitational mass $M=1.4\ms$ of the cold NS with BHF EOS.
The eight parameters of Eqs.~(\ref{e:profile_s},\ref{e:profile_y})
and an estimated initial time $t$
are tabulated for each profile.
}
\begin{ruledtabular}
\begin{tabular}{ccllllllllr}
Label&$M/\ms$&$\smin$&$\smax$& $a$ & $b$ & $h$ & $k$ & $Y_c$&$Y_s$ & $t$ (s) \\
\hline
P1 & 1.47 & 0.6 & 3.6 & 0.35 & 0.134 & 1.0 & 0.451 & 0.38 & 0.08 &  2 \\
P2 & 1.46 & 0.6 & 3.3 & 0.43 & 0.135 & 0.9 & 0.450 & 0.36 & 0.08 &  3 \\
P3 & 1.46 & 1.0 & 2.8 & 0.55 & 0.159 & 0.6 & 0.226 & 0.34 & 0.07 &  5 \\
P4 & 1.45 & 1.0 & 2.6 & 0.65 & 0.159 & 0.5 & 0.113 & 0.33 & 0.07 &  7 \\
P5 & 1.44 & 1.0 & 2.5 & 0.85 & 0.275 & 0.3 & 0.023 & 0.25 & 0.07 & 10 \\
P6 & 1.42 & 1.0 & 2.1 & 1.10 & 0.271 & 0.2 & 0.012 & 0.20 & 0.06 & 15 \\
P7 & 1.41 & 0.6 & 1.6 & 1.30 & 0.273 & 0.1 & 0.002 & 0.16 & 0.05 & 25 \\
P8 & 1.40 & 0.0 & 0.0 &      &       & 0.0 &       &\myc{x_\nu=0}&    \\
\end{tabular}
\end{ruledtabular}
\label{t:profile}
\end{table}

The profiles in Table~\ref{t:profile}
are visualized as functions of the enclosed baryonic mass $m_B$
for the BHF EOS
in Fig.~\ref{f:frad},
where entropy per particle $s$,
corresponding temperature $T$,
lepton fraction $Y_e$, and corresponding
neutrino fraction $x_\nu$
are reported.
The different colors indicate the different time snapshots
from the beginning at $t\sim2\,$s (yellow),
until the cold NS (black).
The figure reveals an initial entropy peak located in the outer-core region,
accompanied by a corresponding temperature maximum
that is fueled by an elevated neutrino fraction.
As neutrino-driven heating progresses,
both the entropy and temperature maxima undergo inward migration
towards the stellar center.
This thermal evolution continues until the central entropy/temperature
approximately doubles its initial value,
marking the transition to the cooling phase.
During this subsequent stage,
the star experiences gradual temperature reduction,
entropy gradient relaxation,
and diminishing neutrino content,
ultimately reaching a neutrino-transparent configuration.
This evolutionary process exhibits remarkable consistency
with the scenario presented in \cite{Pons99,Roberts17}.
It can also be deduced that the constant Y0.4 choice analyzed before
provides an upper limit on the realistic trapping conditions,
even for the earliest stages.

\begin{figure}[t]
\vskip-1mm
\centerline{\includegraphics[width=0.5\textwidth]{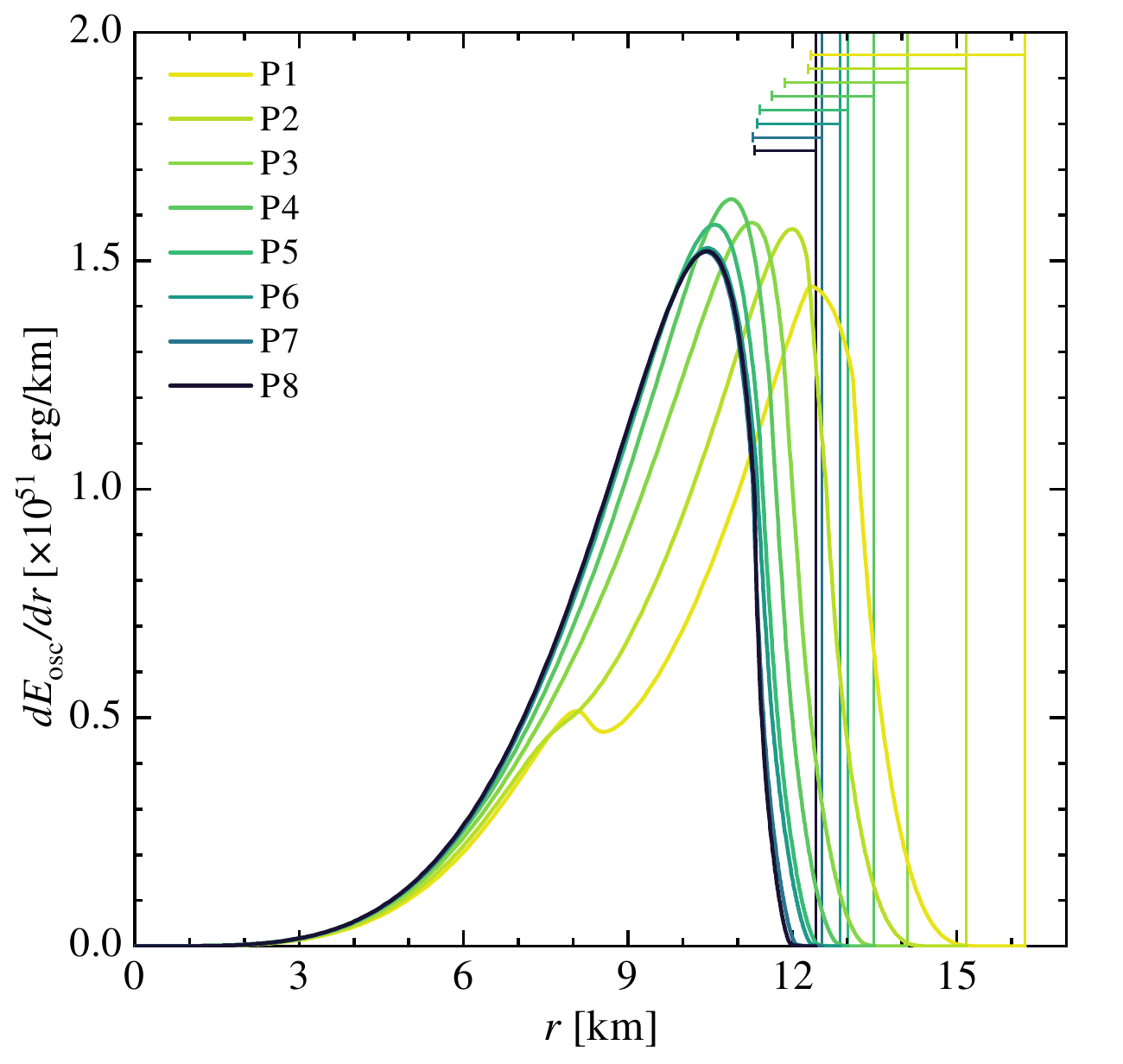}}
\vskip-4mm
\caption{
The radial distribution of oscillation energy,
Eq.~(\ref{e:dedr}),
for $M_B=1.55\ms$ with BHF EOS
for several time snapshots labelled by different colors.
The vertical lines indicate the radius $R$
and the horizontal bars the extension of the crust.
}
\label{f:dedr2}
\end{figure}

\begin{table*}[t]
\def\myc#1{\multicolumn{2}{c}{$#1$}}
\caption{
The gravitational mass $M$,
radius $R$, MOI $I$, tidal deformability $\Lambda$,
frequency $f$ and damping time $\tau$
of the $f$-mode oscillation,
as well as deviations from the
$\Omega_f(\bar{I})$ and $\Omega_f(\Lambda)$ URs
for each profile in Table~\ref{t:profile}.
}
\begin{ruledtabular}
\begin{tabular}{cccccccrrrr}
 &\multicolumn{4}{c}{Macroscopic features}
 &\multicolumn{2}{c}{$f$ mode}
 &\multicolumn{4}{c}{Deviations from fits in \%} \\
 \cline{2-5} \cline{6-7} \cline{8-11}  &&&&&&&
 \myc{ \text{Re}(\text{Im})\Omega_f - \bar{I}} &
 \myc{ \text{Re}(\text{Im})\Omega_f - \Lambda} \\
 Label & $M/\!\ms$ & $R$ [km] & $I$ [km$^3$] & $\Lambda$ & $f$ [kHz] & $\tau$ [s]
 & Zhao fit & Chirenti fit & Zhao fit & Sotani fit \\
\hline
P1 & 1.47 & 16.24 & 140.2 & 651 & 1.56 & 0.29 &-0.94 (-4.62) &-1.49 (-4.53) & 0.03 (-3.65) &-0.16 (-3.12)\\
P2 & 1.46 & 15.18 & 135.0 & 532 & 1.62 & 0.26 &-0.15 (-1.07) &-0.70 (-0.93) &-0.94 (-3.45) &-1.15 (-3.03)\\
P3 & 1.46 & 14.09 & 127.8 & 452 & 1.69 & 0.24 & 0.31 ( 0.81) &-0.22 ( 1.04) &-0.57 (-1.69) &-0.79 (-1.34)\\
P4 & 1.45 & 13.49 & 123.5 & 410 & 1.73 & 0.23 & 0.48 ( 1.41) &-0.05 ( 1.69) &-0.60 (-1.43) &-0.82 (-1.12)\\
P5 & 1.44 & 13.01 & 118.4 & 399 & 1.76 & 0.22 & 0.57 ( 1.72) & 0.04 ( 2.01) &-0.73 (-1.56) &-0.95 (-1.25)\\
P6 & 1.42 & 12.87 & 116.2 & 407 & 1.77 & 0.23 & 0.60 ( 1.81) & 0.07 ( 2.08) &-0.82 (-1.72) &-1.05 (-1.40)\\
P7 & 1.41 & 12.55 & 113.3 & 405 & 1.78 & 0.22 & 0.63 ( 1.93) & 0.10 ( 2.19) &-1.33 (-2.64) &-1.55 (-2.33)\\
P8 & 1.40 & 12.43 & 112.4 & 436 & 1.78 & 0.22 & 0.63 ( 1.94) & 0.10 ( 2.20) & 0.22 ( 0.34) &-0.01 (-0.69)\\
\end{tabular}
\end{ruledtabular}
\label{t:profile2}
\end{table*}

In Fig.~\ref{f:dedr2},
we show the oscillation energy density distributions,
with normalization determined by the boundary condition $W(r=0)=1$
in Eq.~(\ref{e:xi}),
in the PNS at different time snapshots,
for the same profiles as before.
It can be seen that
the star contracts steadily,
accompanied by a gradual inward shift
of the peak position of the oscillatory energy,
which remains located close to the core-crust interface.
This feature is the same as observed for the simplified profiles
used in Fig.~\ref{f:dedr} before.

In Table~\ref{t:profile2} we list for each profile in Table~\ref{t:profile}
the gravitational mass $M$, radius $R$, MOI $I$, tidal deformability $\Lambda$,
frequency $f$ and damping time $\tau$ of the $f$-mode oscillation,
as well as the deviations from the $\Omega_f(\bar{I})$
and $\Omega_f(\Lambda)$ URs.
It can be seen that during the evolution occurs
a monotonic decrease in gravitational mass,
resulting from the combined energy losses
due to neutrino emission and thermal radiation.
The stellar radius progressively contracts from $16.24\km$ to $12.43\km$,
and the MOI decreases accordingly.
In contrast,
the tidal deformability exhibits a gradual decrease
from 2 to 10 seconds after the bounce,
followed by slight rise to the cold-NS value,
consistent with the nonmonotonic temperature dependence
of the results presented in Fig.~\ref{f:tdmoi-m}.

During the evolution we also observe a steady
increase of the $f$-mode frequency from 1.56 to 1.78$\khz$,
and a decrease of the damping time from 0.29 to 0.22 seconds,
until the formation of a cold NS,
both consistent with the results of Fig.~\ref{f:ftau-m}.
During the earlier postbounce phase ($\lesssim 2\,$s),
the oscillation frequency is much lower
(even less than $1\khz$),
as demonstrated in \cite{Torres-Forne19,Mori23,Guedes24,Sotani24}.
This frequency shift places the dominant $f$-mode GW emission
within the optimal sensitivity region of current GW detectors,
significantly enhancing detection prospects.
The evolution of the $f$-mode frequency and damping time resembles the results
presented in \cite{Camelio17},
which lists values ranging from 0.2 to 20 seconds after core bounce.

The last columns of the table examine the validity of the URs
between the $f$-mode characteristics and macroscopic properties
for each profile during the PNS evolution.
The URs are valid within about 2\% (5\%)
for the real (imaginary) parts of $\Omega_f(\bar{I})$
and 2\% (4\%) for $\Omega_f(\Lambda)$,
similar to Figs.~\ref{f:mf-moi/m3},\ref{f:mf-lmbda} for $M=1.4\ms$.
Recent work \cite{Guedes24} has also established that
the $\bar{I}-\Lambda-\bar{Q}$ and $\Omega_f-\Lambda$ URs
remain approximately valid for PNSs during the early postbounce phase.
Their analysis reveals that these relations hold with small deviations
$(\lesssim10\%)$ from roughly 0.5 to 1 second after core bounce,
demonstrating the robustness of these URs even under the extreme thermal
and compositional conditions characteristic of PNSs.
Building upon these results,
our analysis suggests that both URs may remain valid
throughout much of the evolutionary timeline of the PNS,
potentially extending from the early postbounce phase ($\gtrsim 0.5\,$s)
to the subsequent cooling and eventual formation of a cold NS.

\begin{figure}[t]
\vskip-4mm
\centerline{\includegraphics[width=0.5\textwidth]{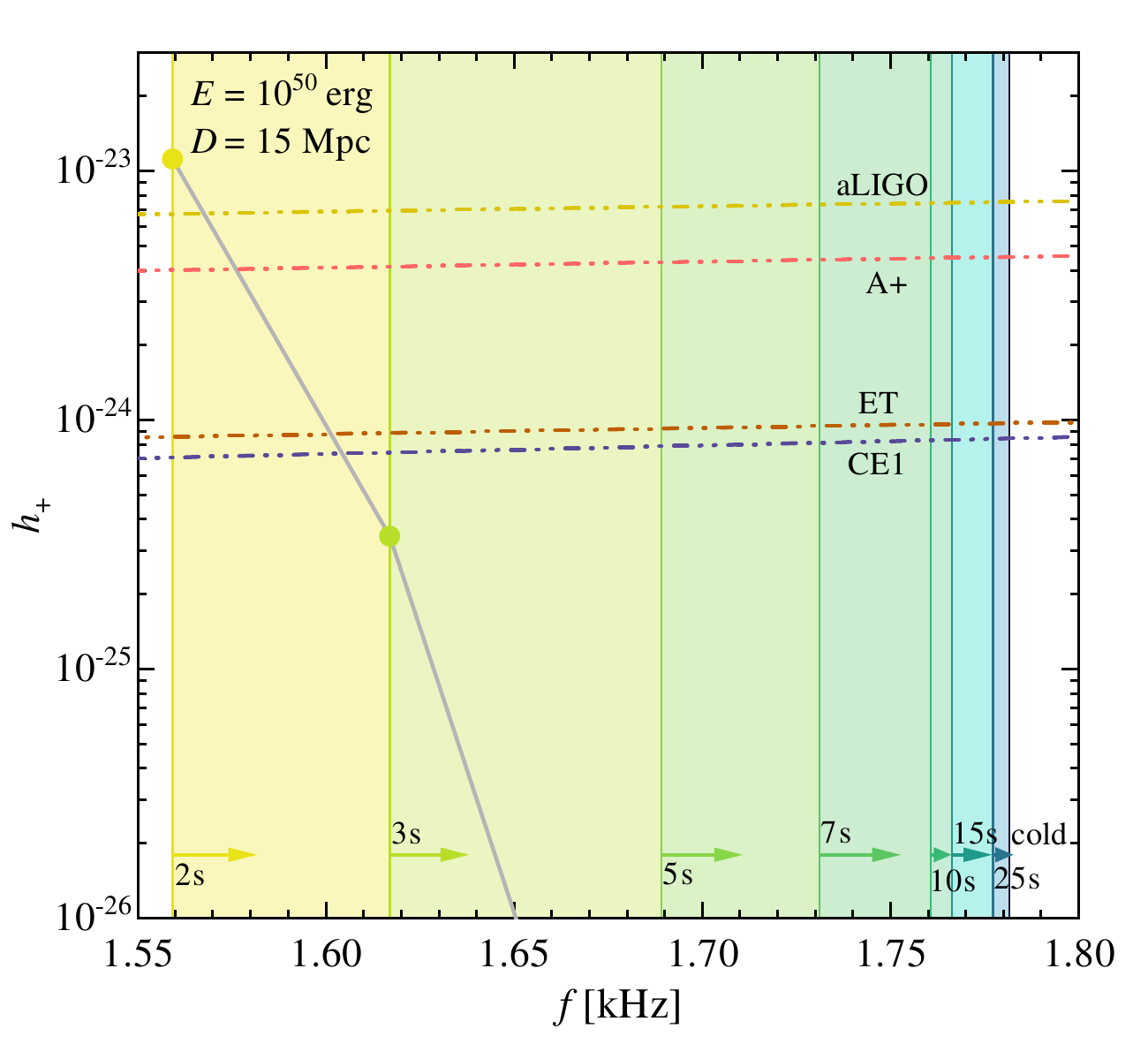}}
\vskip-3mm
\caption{
The GW strain amplitude $h_+$,
Eq.~(\ref{e:hplus}),
of $f$-mode oscillation at frequency $f$,
for a total GW energy $E=10^{50}\;$erg and at distance $D=15\;$Mpc,
obtained with the profiles of Table~\ref{t:profile}.
The estimated initial times are indicated.
The double-dotted lines are sensitivities of GW detectors.
}
\label{f:h-f2}
\end{figure}

Finally,
regarding the detectability of such events,
we assume that the $f$-mode oscillation
is triggered at a time $t_0$ shortly after the core bounce,
with a total GW energy $E=10^{50}\;$erg and at a distance $D=15\;$Mpc.
Subsequently the radiation power $P_\text{GW}$
decays exponentially $\sim \exp[-2(t-t_0)/\tau_i]$
with the damping times $\tau_i$ in Table~\ref{t:profile2}.
In Fig.~\ref{f:h-f2},
we present the corresponding evolution
$\sim \exp[-(t-t_0)/\tau_i]$
of the GW strain amplitude,
Eq.~(\ref{e:hplus}),
assuming $t_0=2\,$s for simplicity.
The amplitude reaches $\sim10^{-23}$ initially,
but then decays rapidly
with the damping time $\tau_i$.
It therefore becomes undetectable
by current and future ground-based GW detectors within less than 1 second.
After 3 seconds,
the strain amplitude has decayed by more than four orders of magnitude,
rendering it entirely undetectable.

\section{Conclusions}
\label{s:end}

PNSs represent hot and lepton-rich compact objects
that exhibit different properties compared to cold, neutrino-transparent NSs.
These thermal and compositional differences
lead to distinct structural characteristics.
We constructed models of PNSs using a finite-temperature EOS
described by the BHF or RMF theory.
Assuming isentropy and fixed lepton fractions for the internal structure,
we then computed the macroscopic features of these stars,
i.e., gravitational mass, radius, MOI, and tidal deformability.
Based on the equilibrium structure,
we solved the oscillation equations within full GR,
and obtained the complex quadrupole $f$-mode eigenfrequencies
for the various EOSs.

The finite-temperature and neutrino-trapping effects influence
both the MOI and tidal deformability of PNSs,
exhibiting strong correlations with
the $f$-mode frequencies and damping times.
This demonstrates the tight connection between $f$-mode characteristics
and the macroscopic properties of (P)NSs.
Previous works have studied $f$-mode frequencies and damping times,
and proposed EOS-insensitive URs with dimensionless MOI $\bar{I}$
and dimensionless tidal deformability $\Lambda$.
The validity for cold NSs has been widely confirmed.
We extended these studies to PNSs
with the assumptions of isentropic conditions and fixed lepton fractions,
showing that the URs are also valid in this case
for not too large neutrino trapping.
The URs were found to be valid within a few percents
for PNSs with gravitational masses exceeding $1.4\ms$.

With the next generation of GW telescopes,
the GWs from quasi-normal modes of NS oscillations may be detectable,
as shown in our previous work \cite{Zheng23,Zheng25}.
In this work,
we demonstrate the detectability of $f$-mode GW radiation from PNSs.
Compared to cold NSs,
the substantially larger energy reservoir available in PNSs
for exciting oscillations,
combined with the enhanced probability of oscillation-triggering mechanisms
during the early evolutionary stages and the lower frequencies of GWs,
may effectively increase the probability of detection.

According to simulation results,
we then modeled details of the thermal and trapping radial profiles
by using the BHF EOS for a PNS with canonical mass
from about $2\,$s after core bounce to the formation of a cold NS.
During this process,
the $f$-mode frequency gradually increases from $1.56\khz$ to $1.78\khz$,
and the URs remain valid within a few percent.
Assuming a total GW energy $E=10^{50}\;$erg and a distance $D=15\;$Mpc,
the initial GW strain amplitude $\sim10^{-23}$
would be detectable during the timescale of the damping time $\tau\sim0.3\,$s
by current and future GW detectors.

This study can be extended to include non-nucleonic degrees of freedom
such as hyperons or quark matter with a hot EOS.
In addition, the (differential) rotation of the star
and a magnetic field affect the structure of the star,
as well as the oscillations.
We leave these issues to future work.

\begin{acknowledgments}

Zi-Yue Zheng and Xiao-Ping Zheng are supported by
the National Natural Science Foundation of China (Grant No.~12033001)
and the National SKA Program of China (Grant No.~2020SKA0120300).
Jin-Biao Wei, Huan Chen, and Ting-Ting Sun acknowledge financial support
from the National Natural Science Foundation of China (Grant No.~12205260).

\end{acknowledgments}

\newcommand{\epja}{Euro. Phys. J. A}
\newcommand{\aap}{Astron. Astrophys.}
\newcommand{\apjl}{Astrophys. J. Lett.}
\def\jcap{Journal of Cosmology and Astroparticle Physics}
\def\jcap{JCAP}
\newcommand{\mnras}{Mon. Not. R. Astron. Soc.}
\newcommand{\nphysa}{Nucl. Phys. A}
\newcommand{\physrep}{Phys. Rep.}
\newcommand{\plb}{Phys. Lett. B}
\newcommand{\ppnp}{Prog. Part. Nucl. Phys.}
\bibliographystyle{apsrev4-1}
\bibliography{pnsnro}

\end{document}